\documentclass[journal,10pt]{IEEEtran} 
\makeatletter
\def\endthebibliography{%
	\def\@noitemerr{\@latex@warning{Empty `thebibliography' environment}}%
	\endlist
}
\makeatother

\usepackage{cite}

\usepackage[pdftex]{graphicx}

\usepackage[caption=false,font=footnotesize]{subfig}

\usepackage{amsmath}
\usepackage{mathtools, cuted}
\usepackage{pifont}
\usepackage{filecontents}
\usepackage{amssymb}
\usepackage{color}
\usepackage{xcolor}
\usepackage{booktabs}

\usepackage{epsfig}
\usepackage{bm}

\usepackage{url}

\usepackage{algpseudocode}
\usepackage{algorithm, tabularx}

\usepackage{lettrine}

\usepackage{lipsum}

\usepackage{siunitx}

\usepackage{soul,color}

\usepackage{mathrsfs}

\usepackage{array}

\usepackage[inline]{enumitem}

\usepackage{breqn}

\usepackage{multirow}
\usepackage{array}
\newcolumntype{L}[1]{>{\raggedright\let\newline\\\arraybackslash\hspace{0pt}}m{#1}}
\newcolumntype{C}[1]{>{\centering\let\newline\\\arraybackslash\hspace{0pt}}m{#1}}
\newcolumntype{R}[1]{>{\raggedleft\let\newline\\\arraybackslash\hspace{0pt}}m{#1}}

\usepackage{amsthm}
\newtheorem{theorem}{Theorem}

\newtheorem{lemma}[theorem]{Lemma}

\newtheorem{corollary}{Corollary}[theorem]

\newlength{\maxwidth}
\newcommand{\algalign}[2]
{\makebox[\maxwidth][r]{$#1{}$}${}#2$}

\makeatletter
\newcommand{\multiline}[1]{%
	\begin{tabularx}{\dimexpr\linewidth-\ALG@thistlm}[t]{@{}X@{}}
		#1
	\end{tabularx}
}
\makeatother

\DeclareMathOperator*{\argmax}{\arg\max}
\theoremstyle{remark}

\newtheorem{remark}{Remark}

\algdef{SE}[SUBALG]{Indent}{EndIndent}{}{\algorithmicend\ }%
\algtext*{Indent}
\algtext*{EndIndent}

\newcommand{\cmark}{\ding{51}}%

\begin{document}
	
	\title{FRIS-Empowered Wireless Networks: Joint Port-Phase Design for Outage-Capacity Synergy}
		
	\author{Hong-Bae Jeon,~\IEEEmembership{Member,~IEEE}, Tuo Wu, Kai-Kit Wong,~\IEEEmembership{Fellow,~IEEE}, and Chan-Byoung Chae,~\IEEEmembership{Fellow,~IEEE}
		
		\thanks{\textit{(Corresponding Author: Chan-Byoung Chae)}}
		\thanks{H.-B. Jeon is with the School of Electronic Engineering, Soongsil University, Seoul, Korea (e-mail: hongbae08@ssu.ac.kr).}
		\thanks{T. Wu is with the School of Electronic and Information Engineering, South China University of Technology, Guangzhou, China (e-mail:wutuo@scut.edu.cn).}
		\thanks{K.-K. Wong is with the Department of Electronic and Electrical Engineering, University College London, WC1E 6BT London, U.K., and also with the Yonsei Frontier Laboratory, Yonsei University, Seoul, Korea (e-mail: kai-kit.wong@ucl.ac.uk).}
		\thanks{C.-B. Chae is with the School of Integrated Technology, Yonsei University, Seoul, Korea (e-mail: cbchae@yonsei.ac.kr).}
	}
	

	\maketitle

\begin{abstract}
In this paper, we propose two alternating optimization (AO) frameworks that jointly optimize port selection and phase shifts in fluid reconfigurable intelligent surface (FRIS)-aided wireless networks to minimize the outage probability and enhance the ergodic capacity. For port selection, we develop (i) a cross-entropy-method (CEM)-based approach that efficiently explores the combinatorial search space via elite-sample updates, and (ii) a semidefinite relaxation (SDR)-based method that optimizes continuous port weights followed by discrete selection. For phase optimization, we introduce a Dinkelbach-based Riemannian gradient method with penalty on the complex unit-modulus manifold. We prove that any Pareto improvement in the equivalent channel parameters simultaneously reduces the outage probability and increases the ergodic capacity, which motivates the Pareto-based design of the proposed AO frameworks. Simulation results show that the proposed frameworks significantly outperform conventional schemes, with AO-SDR achieving the highest reliability and capacity gains and AO-CEM offering a favorable performance-complexity trade-off.
\end{abstract}


	\IEEEpeerreviewmaketitle
	
	\section{Introduction}
\lettrine{R}{ecently}, the technology of reconfigurable intelligent surface (RIS) has swiftly matured into a central element in sixth-generation (6G) wireless communications~\cite{RISvtm, RIST}. RIS is envisioned as planar programmable metasurfaces composed of numerous low-cost reflecting elements capable of dynamically controlling the wireless propagation environment, enhancing coverage, capacity, energy efficiency, and localization capabilities in modern networks~\cite{holotut}. By adjusting the phase shifts of individual elements, RIS can deliberately redirect incident signals, thereby shaping the propagation environment to enhance a wide range of applications~\cite{dsRIS, HBRIS, tkz}.

Although RIS offers remarkable advantages, its practical deployment is hindered by several intrinsic challenges. Unlocking its full potential requires massive arrays with even thousands of reflecting elements~\cite{risleo, losris}, which inevitably incur considerable hardware costs and impose heavy control burdens~\cite{risholonear}. Furthermore, densely packed configurations give rise to strong spatial correlation~\cite{risspa}, thereby constraining the available spatial degree-of-freedoms (DoFs) and limiting achievable performance~\cite{risexdof, ristitdof}. At the same time, the optimization burden, spanning channel estimation~\cite{CE4} and phase adjustment~\cite{risphase1, risphase2}, grows disproportionately with the number of elements, resulting in prohibitive computational costs for real-world deployments~\cite{rismeasure, rismeasure22}. Furthermore, existing RIS architectures are physically inflexible~\cite{simpleris, flexris}, preventing the full exploitation of spatial diversity even in scenarios where ample physical space is available. Taken together, these challenges underscore the need for novel design strategies that retain the advantages of RIS while mitigating its spatial limitations.

To empower RIS for greater spatial adaptivity, it is possible to leverage the fluid antenna system (FAS) concept for shape and position reconfigurability in antennas~\cite{FAS}. In recent years, FAS has been shown to be effective for many applications~\cite{fluidtut, fasmag}. In~\cite{FASqout}, the authors demonstrate that multiple-input-multiple-output (MIMO)-FAS markedly surpasses conventional MIMO architectures by exploiting adaptive port selection and spatial diversity, introducing the $q$-outage capacity metric to capture these gains. In~\cite{perlimfas}, it is shown that even a compact single-antenna FAS can provide notable capacity improvements, with exact ergodic capacity expressions and tight lower bounds, revealing that such systems can rival the performance of traditional multi-antenna maximum-ratio combining (MRC) setups. To address the challenges posed by the intricate spatial correlation inherent to FAS, the study in~\cite{newcor} proposes a spatial block-correlation model as a tractable yet accurate approximation for arbitrary 1D and 2D FAS correlation patterns, thereby simplifying analysis and simulation compared to classical models like Clarke’s or Jakes'. In~\cite{fasopdg}, the authors derive closed-form results for outage probability and diversity gain, and further introduce a near-optimal port selection algorithm with reduced computational overhead. Taken together, these studies establish FAS as a promising paradigm that surpasses conventional antenna systems by unlocking higher spatial diversity, improving outage and capacity performance, and enabling efficient yet practical analysis for compact wireless devices.
\begin{table*}[t]
\centering
\caption{Comparison of Studies on FRIS}
\label{tab:FRIS_comparison}
\renewcommand{\arraystretch}{1.2}
\begin{tabular}{lcccccc}
\toprule
\textbf{Reference} & \cite{FRISlook} & \cite{FRISonoff} & \cite{FRISpa} & \cite{FRISISAC} & \cite{VegaSanchez2026}& \textbf{This Work} \\
\midrule
Comprehensive Scope of Performance Analysis & -- & -- & \cmark & -- & -- & \cmark  \\
Joint Optimization on FRIS Port-Phase       & \cmark & \cmark & -- & \cmark & \cmark & \cmark  \\
Comprehensive Analysis on Computational Complexity & \cmark & -- & -- & \cmark & \cmark & \cmark  \\
Validation of Optimality                     & -- & -- & -- & -- & -- & \cmark  \\
\bottomrule
\end{tabular}
\end{table*}

Although these reconfigurable antenna architectures such as FAS have expanded the design space well beyond traditional fixed-geometry antennas~\cite{bvcfas, pixfas, trimimo, trimimo22, nfris}, the vast majority of existing efforts remain firmly anchored in cellular link enhancement~\cite{fluidtut, fastnse}. Their potential in spatial and geometric flexibility within RIS-assisted wireless systems remains largely unexplored, which, in turn, motivates the incorporation of fluidic motion control into conventional RIS architectures. Inspired by this concept, the notion of a fluid RIS (FRIS) has been introduced, which proposes that the RIS elements can also dynamically select their spatial positions within a predefined area~\cite{FRISlook, FRISonoff, FRISpa, VegaSanchez2026, frissec, frispat, FRISISAC}. This geometric flexibility allows FRIS to harness greater spatial diversity and improve link reliability under dynamic wireless conditions.

Early works have highlighted different aspects of FRIS. In~\cite{FRISlook}, an initial FRIS framework was proposed and optimized under different antenna configurations and user scenarios, showcasing significant rate improvements enabled by position reconfigurability of FRIS. In~\cite{FRISonoff}, a practical model with on-off element control and discrete phase shifts was studied, where a cross-entropy-method (CEM) based algorithm achieved significant improvements in rate under hardware constraints. Furthermore,~\cite{FRISpa} provided a statistical framework with closed-form approximations for outage probability and ergodic capacity, showing that FRIS can enhance both link reliability and spectral efficiency compared to conventional RIS systems. In~\cite{VegaSanchez2026}, the authors combined evolutionary particle swarm optimization (E-PSO) for FRIS configuration to maximize the received SNR, and analyzed the resulting outage probability under spatially correlated fading. In~\cite{frissec}, the authors investigated FRIS-assisted secure communications and developed a framework to jointly optimize transmit beamforming and FRIS configuration for secrecy-rate maximization. In~\cite{frispat}, the authors proposed a pattern-reconfigurable FRIS architecture with an optimization on element-level radiation patterns to maximize the weighted sum rate. Further applications of FRIS to 6G use-cases such as integrated-sensing-and-communication (ISAC) are presented in~\cite{FRISISAC} with demonstrating its benefits. Despite these initial findings, research on the overall performance characterization and optimization of FRIS-assisted systems remains at an early stage, and comprehensive validation of their optimality and computational complexity has yet to be established.

Motivated by these observations, in this paper we develop a comprehensive framework for the analysis and optimization of FRIS-assisted wireless networks. We establish a port-and-phase selection model for FRIS-aided signal transmission, and derive a unified outage probability characterization that explicitly incorporates spatial correlation among the selected FRIS ports. Building on this, we formulate an optimization problem aimed at minimizing outage probability by jointly determining both the port configuration and phase adjustments of the FRIS. To tackle this problem, we address its inherently non-convex and multi-objective nature by developing an alternating optimization (AO) framework that jointly balances optimization and computational tractability, which also leverages the Pareto-optimal balance of two objectives, while also demonstrating analytically that Pareto improvements reducing the outage probability also improve the ergodic capacity. Through rigorous analysis and extensive simulations, we deliver practical design guidelines and key insights to enable efficient port-and-phase selection in FRIS systems. The main contributions are summarized as follows:
\begin{enumerate}
\item We establish a joint port selection and phase design framework for FRIS-assisted wireless networks, which explicitly accounts for the spatial correlation among the selected ports and yields a unified characterization of the outage probability. Based on this, we formulate an outage minimization problem over the port configuration and phase shifts.
\item Since the outage probability depends on two configuration-dependent channel parameters that are difficult to improve simultaneously, we formulate a Pareto-based surrogate problem and develop two AO frameworks. For port selection, we propose (i) a CEM-based method with elite-sample updates and (ii) a semidefinite relaxation (SDR)-based method that optimizes continuous port weights followed by discrete selection. For phase optimization, we propose a Dinkelbach-based Riemannian gradient method with penalty on the complex unit-modulus manifold.
\item We show that any Pareto improvement of the two channel parameters reduces the outage probability at every target rate and increases the ergodic capacity, revealing the complementary relationship between reliability and capacity in FRIS systems. We also analyze the computational complexity.
\item Extensive Monte Carlo simulations over various environments show that the proposed SDR-based framework achieves the highest reliability and capacity, whereas the proposed CEM-based one attains comparable performance at substantially lower complexity. Moreover, both frameworks achieve near-optimal performance compared with brute-force search (BFS) over discrete phase configurations, at a fraction of its runtime.
\end{enumerate}
\begin{figure}[t]
	\begin{center}
		\includegraphics[width=0.95\columnwidth,keepaspectratio]%
		{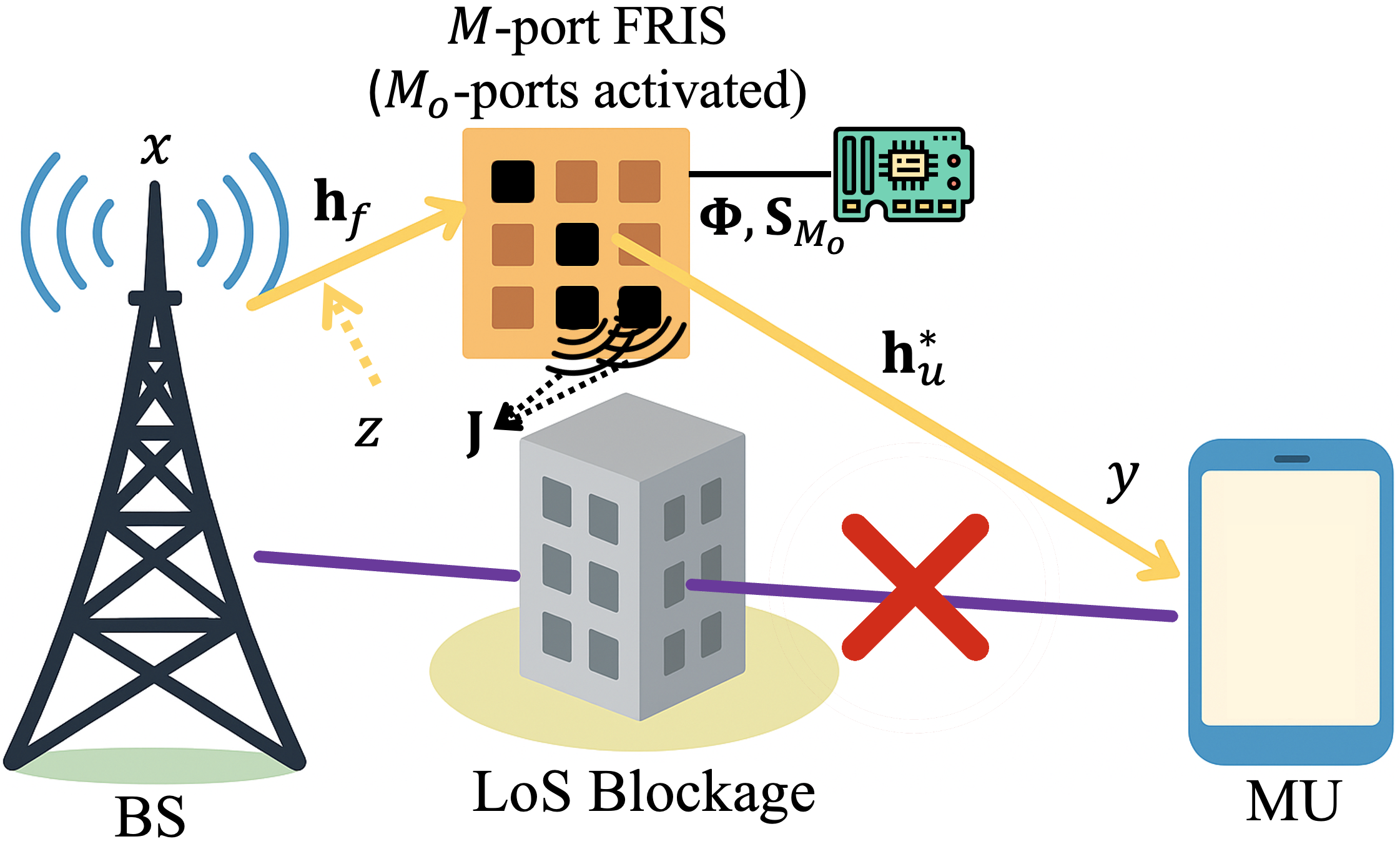}
		\caption{System model of the FRIS-aided wireless network.}
		\label{fig_overview}
	\end{center}
\end{figure}
\section{System Model}\label{SM2}
As illustrated in Fig.~\ref{fig_overview}, we consider a downlink wireless communication scenario assisted by a FRIS. A base station (BS) equipped with a single fixed-position antenna (FPA) communicates with a mobile user (MU), also equipped with a single FPA, via the FRIS, under the assumption that the direct BS-MU path is completely obstructed by obstacles such as buildings. The FRIS comprises $M = M_x \times M_x$ reflecting elements deployed over a surface of dimensions $W_x\lambda_c \times W_x\lambda_c$, where $\lambda_c$ denotes the carrier wavelength and $W_x$ represents the normalized aperture of the FRIS relative to $\lambda_c$. The elements are densely spaced with an inter-element distance $d = \frac{W_x \lambda_c}{M_x}$, resulting in a spatial correlation $\mathbf{J} \triangleq [J_{ij}] \in \mathbb{R}^{M \times M}$; the spatial correlation $J_{ij}$ of the $i$ and $j$th element is~\cite{FRISonoff, FRISpa, FASqout} $J_{ij} = j_0\left( \frac{2\pi d_{ij}}{\lambda_c} \right)$, where $j_0$ denotes the zero-order spherical Bessel function of the first kind, and $d_{ij}$ is the distance between the $i$ and $j$th port ($i,j\in\{1,\cdots,M\}$):
\begin{equation}
\label{dij}
d_{ij}= \|\mathbf{p}_i-\mathbf{p}_j\|_2,~\mathbf{p}_i\triangleq d\Big[\operatorname{mod}(i-1,M_x),~\Big\lfloor\frac{i-1}{M_x}\Big\rfloor\Big]^{\mathrm{T}},
\end{equation}
with $\mathbf{p}_i$ denoting the position of the $i$th port on the FRIS plane. 
Compared to~\cite{FRISlook} and in line with~\cite{FRISonoff}, each FRIS element in our model can be interpreted as a port within a fluid element of FAS. Therein, each element supports two states: `on' and `off'. In the `on' state, the element interacts with the impinging electromagnetic wave, altering its properties to realize the intended system function. In the `off' state, the element is terminated with a matched load, thereby isolating it from the incident wave and preventing any signal modification. 

Let $M_o$ denote the number of active (`on') elements, and define the corresponding selection matrix as $\mathbf{S}_{M_o}^{\mathrm{T}} \triangleq [\mathbf{e}_{i_1} \cdots \mathbf{e}_{i_{M_o}}] \in \mathbb{R}^{M \times M_o}$ for $\forall i_n \in \{1, \cdots, M\}$, where $\mathbf{e}_i$ is the $i$th column of the $M \times M$ identity matrix $\mathbf{I}_M$. This implies that the $\{i_n\}_{n=1}^{M_o}$th ports of the FRIS are activated. To avoid redundant selection, any two columns of $\mathbf{S}_{M_o}^{\mathrm{T}}$ must be distinct. Equivalently, the activation state of the ports is described by $\boldsymbol{\chi}\in\{0,1\}^M$ with $\mathbf{1}^{\mathrm{T}}\boldsymbol{\chi}=M_o$ (1: on, 0: off), so that $\mathbf{S}_{M_o}^{\mathrm{T}}\mathbf{S}_{M_o}=\mathrm{diag}(\boldsymbol{\chi})$. To tie each phase shift to a physical port, we adopt the port-indexed reflection matrix $\boldsymbol{\Phi} \triangleq \mathrm{diag}(\{e^{j\phi_m}\}_{m=1}^{M})\in\mathbb{C}^{M\times M}$ with $\forall \phi_m \in [0, 2\pi)$, where $\phi_m$ is the phase shift of the $m$th port. Only the entries of the active ports, i.e., $\mathbf{S}_{M_o}\boldsymbol{\Phi}\mathbf{S}_{M_o}^{\mathrm{T}}=\mathrm{diag}(\{e^{j\phi_{i_n}}\}_{n=1}^{M_o})$, affect the signal, and the inactive ports carry the default unit gain ($\phi_m=0$). The overall FRIS operation is then represented by:
\begin{equation}
\label{defa}
\mathbf{A} \triangleq \mathbf{J}^{\frac{1}{2}} \mathbf{S}_{M_o}^{\mathrm{T}} \big(\mathbf{S}_{M_o}\boldsymbol{\Phi}\mathbf{S}_{M_o}^{\mathrm{T}}\big) \mathbf{S}_{M_o} \mathbf{J}^{\frac{1}{2}}=\mathbf{J}^{\frac{1}{2}} \mathrm{diag}(\boldsymbol{\chi})\boldsymbol{\Phi} \mathbf{J}^{\frac{1}{2}},
\end{equation}
where $\mathbf{J}^{\frac{1}{2}}$ denotes the matrix square root of $\mathbf{J}$
~\cite{FRISonoff, FRISlook}. This port-indexed parameterization makes $\mathbf{A}$, and hence every performance metric below, a well-defined function of $(\mathbf{S}_{M_o},\boldsymbol{\Phi})$ that is invariant to the ordering of $\{i_n\}_{n=1}^{M_o}$: when the port selection changes, each retained port keeps its own phase shift, whereas a newly activated port starts from unit gain. Following~\cite{FRISonoff, FRISlook}, the spatial correlation among the densely packed ports is captured by applying $\mathbf{J}^{\frac{1}{2}}$ to the overall FRIS response in~(\ref{defa}). Since this operator acts on both the deterministic LoS and the scattered components below, it should be interpreted as an effective port-correlation (spatial-smoothing) operator of the FRIS rather than as a property of the propagation channel itself; it is a modeling abstraction adopted for analytical tractability and consistency with the FRIS literature. As the port spacing $d$ grows, $\mathbf{J}\to\mathbf{I}_M$ and~(\ref{defa}) reduces to the conventional RIS model $\mathbf{A}=\mathrm{diag}(\boldsymbol{\chi})\boldsymbol{\Phi}$, whereas an element-level electromagnetic model of the FRIS response, e.g., including mutual coupling, is left for future work. The FRIS-to-MU channel $\mathbf{h}_u$ is modeled according to a Rician fading distribution~\cite{FASRIS}:
\begin{equation}
\label{frismu}
\mathbf{h}_u = \sqrt{\frac{K}{K+1}} \bar{\mathbf{h}}_u + \sqrt{\frac{1}{K+1}} \tilde{\mathbf{h}}_u\in\mathbb{C}^{M\times 1},
\end{equation}
where $K$ is the Rician factor, $\tilde{\mathbf{h}}_u \sim \mathcal{CN}(0, \mathbf{I}_M)$ and $\bar{\mathbf{h}}_u$ is the deterministic LoS component. We assume that the channel $\mathbf{h}_f$ from BS to FRIS is assumed to be pure LoS: $\mathbf{h}_f = \bar{\mathbf{h}}_f\in\mathbb{C}^{M\times 1}$~\cite{nearris, maxminris, FASRIS, FASRIS22}, which is reasonable when the FRIS is deployed in the vicinity of the BS with a clear LoS link~\cite{nearris}. Throughout, the LoS components are modeled by planar wavefronts, which is accurate when the link distance exceeds the Rayleigh distance of the FRIS aperture, $D^{\mathrm{R}}\triangleq 2D_{\mathrm{F}}^2/\lambda_c=4W_x^2\lambda_c$ with the FRIS diagonal $D_{\mathrm{F}}=\sqrt{2}W_x\lambda_c$~\cite{neardai}. The equivalent channel is therefore given by $\mathbf{H}_{\text{eq}} \triangleq \mathbf{h}_u^* \mathbf{A} \bar{\mathbf{h}}_f$, and the received $y$ is:
\begin{equation}
\label{rxy}
y = \sqrt{P L_f L_u} \mathbf{H}_{\text{eq}} x + v,
\end{equation}
where $x$ denotes the unit-power transmitted symbol, $P$ is the transmit power, $v \sim \mathcal{CN}(0, \sigma_0^2)$ represents the additive noise, and $L_k = \rho l_k^{-\alpha}$ denotes the large-scale path-loss for $k \in \{f, u\}$, corresponding to the BS-FRIS and FRIS-MU links, respectively. Here, $\alpha$ is the path-loss exponent (PLE), $\rho$ is the path-loss at unit distance, and $\{l_k\}$ represents the respective BS-FRIS and FRIS-MU link distances.
\subsection{Statistical Characterization of Outage Probability $P_{\text{out}}$}
The equivalent channel gain $|\mathbf{H}_{\text{eq}}|^2$ is expressed as $|\mathbf{H}_{\text{eq}}|^2 = |\mathbf{h}_u^* \mathbf{A} \bar{\mathbf{h}}_f|^2$. Given that $\mathbf{h}_u$ follows a Rician distribution and $\bar{\mathbf{h}}_f$ is deterministic, we expand $\mathbf{H}_{\text{eq}}$ as $\mathbf{H}_{\text{eq}} = \left( \sqrt{\frac{K}{K+1}} \bar{\mathbf{h}}_u^* + \sqrt{\frac{1}{K+1}} \tilde{\mathbf{h}}_u^* \right) \mathbf{A} \bar{\mathbf{h}}_f$, which follows $\mathbf{H}_{\text{eq}}\sim\mathcal{CN}(\mu, \sigma^2 )$ with:
\begin{equation}
\label{meanva}
\mu = \sqrt{\frac{K}{K+1}} \bar{\mathbf{h}}_u^* \mathbf{A} \bar{\mathbf{h}}_f,~\sigma^2 = \frac{1}{K+1} \|\mathbf{A} \bar{\mathbf{h}}_f\|_2^2,
\end{equation}
so that $\frac{2}{\sigma^2}|\mathbf{H}_{\text{eq}}|^2$ follows the non-central chi-squared distribution with DoF 2 and non-centrality parameter $2\lambda$; $\lambda \triangleq \frac{|\mu|^2}{\sigma^2}= K\frac{ \left| \bar{\mathbf{h}}_u^* \mathbf{A} \bar{\mathbf{h}}_f \right|^2 }{ \| \mathbf{A} \bar{\mathbf{h}}_f \|_2^2 }$. Thus, the outage probability $P_{\text{out}}$ is~\cite{FAS}:
\begin{equation}
\label{outr}
\begin{aligned}
P_{\text{out}} &\triangleq \mathbb{P} \left( \log_2(1 + \gamma L_f L_u |\mathbf{H}_{\text{eq}}|^2) \le R \right)\\
& = 1 - Q_1 \left(\sqrt{2\lambda},\sqrt{ \frac{2(2^R - 1)}{\gamma L_f L_u\sigma^2}}\right).
\end{aligned}
\end{equation}
where $R$ and $\gamma\triangleq \frac{P}{\sigma_0^2}$ is the target rate and the transmit SNR, respectively, and $Q_1$ is the first-order Marcum-$Q$ function. By substituting $\lambda$ and $\sigma^2$ to~(\ref{outr}), $P_{\text{out}}$ becomes
\begin{equation}
\label{outageprob}
P_{\text{out}} = 1 - Q_1 \left(
\sqrt{ \frac{2K \left| \bar{\mathbf{h}}_u^* \mathbf{A} \bar{\mathbf{h}}_f \right|^2 }{ \| \mathbf{A} \bar{\mathbf{h}}_f \|_2^2 } },\
\sqrt{ \frac{2(K+1)(2^R - 1)}{ \gamma L_f L_u \| \mathbf{A} \bar{\mathbf{h}}_f \|_2^2 } }\right).
\end{equation}
\subsection{Outage Probability Minimization Problem}
Our goal is to jointly optimize $\mathbf{S}_{M_o}$ and $\boldsymbol{\Phi}$ to minimize $P_{\text{out}}$. The resulting optimization problem is formulated as:
\begin{equation}
\begin{aligned}
\max_{\mathbf{S}_{M_o}, \boldsymbol{\Phi}} & ~ \mathcal{J}(\mathbf{S}_{M_o}, \boldsymbol{\Phi}) \\
&\triangleq Q_1 \left(
\sqrt{ \frac{2K \left| \bar{\mathbf{h}}_u^* \mathbf{A} \bar{\mathbf{h}}_f \right|^2 }{ \| \mathbf{A} \bar{\mathbf{h}}_f \|_2^2 } }, \sqrt{ \frac{2(K+1)(2^R - 1)}{ \gamma L_f L_u \| \mathbf{A} \bar{\mathbf{h}}_f \|_2^2 } }
\right) \\
\text{s.t.} ~ & \mathbf{A} \triangleq \mathbf{J}^{\frac{1}{2}} \mathbf{S}_{M_o}^{\mathrm{T}}\mathbf{S}_{M_o} \boldsymbol{\Phi} \mathbf{J}^{\frac{1}{2}},~\mathbf{S}_{M_o} \in \mathcal{P}_{M_o}, \\
& \boldsymbol{\Phi} \triangleq \mathrm{diag}(\{e^{j\phi_m}\}_{m=1}^{M})~(\forall \phi_m \in [0, 2\pi)),
\end{aligned}
\label{objoriginal}
\end{equation}
where $\mathcal{P}_{M_o}\subset\mathbb{R}^{M_o \times M}$ denotes the set of all possible $M_o$-element selection matrices from $M$ total elements. Problem~(\ref{objoriginal}) is highly intricate since the discrete variable $\mathbf{S}_{M_o}$ and the continuous variable $\boldsymbol{\Phi}$ are jointly involved in both the objective and the constraints, making optimization extremely intractable. 
For later use, we write the two configuration-dependent parameters in~(\ref{meanva}) explicitly as
\begin{equation}
\label{lamsig}
\lambda(\mathbf{S}_{M_o},\boldsymbol{\Phi})\triangleq K\frac{|\bar{\mathbf{h}}_u^*\mathbf{A}\bar{\mathbf{h}}_f|^2}{\|\mathbf{A}\bar{\mathbf{h}}_f\|_2^2},~\sigma^2(\mathbf{S}_{M_o},\boldsymbol{\Phi})\triangleq\frac{\|\mathbf{A}\bar{\mathbf{h}}_f\|_2^2}{K+1},
\end{equation}
so that $\mathcal{J}(\mathbf{S}_{M_o},\boldsymbol{\Phi})=Q_1\big(\sqrt{2\lambda},\sqrt{c/\sigma^2}\big)$ with $c\triangleq\frac{2(2^R-1)}{\gamma L_fL_u}>0$. Since $Q_1(a,b)$ is increasing in $a$ and decreasing in $b$~\cite{marQ}, $\mathcal{J}$ is nondecreasing in each of $\lambda$ and $\sigma^2$. However, $\lambda$ and $\sigma^2$ are coupled through $(\mathbf{S}_{M_o},\boldsymbol{\Phi})$ and generally cannot be maximized simultaneously, which introduces a Pareto-type trade-off between them~\cite{multio}.
\section{Proposed Alternating Optimization}
To effectively address the problem, we develop AO frameworks that iteratively update $\mathbf{S}_{M_o}$ and $\boldsymbol{\Phi}$ in an alternating manner until convergence is achieved.
\subsection{Optimizing $\mathbf{S}_{M_o}$: 1. CEM for FRIS Port Selection}\label{seccem}
Given $\boldsymbol{\Phi}$, we aim to optimize $\mathbf{S}_{M_o}$ by solving the following port selection derived from~(\ref{objoriginal}):
\begin{equation}
\label{cemobj}
\begin{aligned}
\max_{\mathbf{S}_{M_o}\in \mathcal{P}_{M_o} } ~ \mathcal{J}(\mathbf{S}_{M_o}, \boldsymbol{\Phi}).
\end{aligned}
\end{equation}
To tackle the combinatorial nature of the selection problem, we employ the CEM~\cite{FRISonoff, fdris, CEM}. Let \(\chi_m\) denote the activation state of the \(m\)th candidate port. By denoting the scalars represent the individual elements of the respective matrix, we maintain a distribution $\mathbf{g}^{(t)} \in [0,1]^M$, where $g_m^{(t)}$ represents the selection probability of the $m$th FRIS port at iteration $t$. For the initial iteration, $\mathbf{g}^{(0)}$ is obtained from the distribution $\mathbb{P}_{\mathbf{g}^{\star}}$ with optimum $\mathbf{g}^{\star}$ in the previous AO iteration; if it is the first AO iteration, $\mathbf{g}^{\star}$ is initialized as $\mathbf{g}^{\star} \sim \mathcal{U}[0, 1]^M$, where $\mathcal{U}[\cdot]$ denotes the uniform distribution.

At each iteration, we sample $N_{\mathrm{s}}$ binary port selection vectors $\boldsymbol{\chi}^{(s)} \in \{0,1\}^M~(s=1, \cdots, N_{\mathrm{s}})$ (1: on, 0: off) according to the product of Bernoulli distributions:
\begin{equation}
\label{bernoullisamp}
\mathbb{P}_{\mathbf{g}^{(t)}}(\boldsymbol{\chi}) 
= \prod_{m=1}^M \left( g_m^{(t)\chi_m} (1 - g_m^{(t)})^{1 - \chi_m} \right),
\end{equation}
that is, $\mathbb{P}[\chi_m=1]=g_m^{(t)}~(\forall m)$. 
To ensure \( \boldsymbol{\chi} \in \{0,1\}^M \) satisfies $\sum_{m=1}^{M} \chi_m = M_o$, we perform the following correction procedure based on the final selection probability vector $\mathbf{g}^{(t)}$:
\begin{enumerate}
    \item ${\sum_{m=1}^{M} \chi_m > M_o}$: Let $\mathcal{I}_1 \triangleq \{ m : \chi_m = 1 \}$ denote the set of selected indices. Sort \( \mathcal{I}_1 \) in ascending order of \( g_m^{(t)} \), and flip the entries of \( \chi_m \) from 1 to 0 for the smallest \( |\mathcal{I}_1| - M_o \) elements; ensures $ \sum_{m=1}^{M} \chi_m = M_o$.
    \item ${\sum_{m=1}^{M} \chi_m < M_o}$: Let \( \mathcal{I}_0 \triangleq \{ m : \chi_m = 0 \} \) denote the unselected indices set. Sort \( \mathcal{I}_0 \) in descending order of \( g_m^{(t)} \), and flip the entries of $ \chi_m $ from 0 to 1 for the largest \( M_o - \sum_{m=1}^{M} \chi_m \) elements; ensures $ \sum_{m=1}^{M} \chi_m = M_o$.
\end{enumerate}
After this correction, $\boldsymbol{\chi}$ 
is used to construct $ \mathbf{S}_{M_o}^{\star} $.

Let \(\mathcal{E}^{(t)}\) denote the elite set consisting of the top-$\kappa\%$ samples of $\{\boldsymbol{\chi}^{(s)}\}_{s=1}^{N_{\mathrm{s}}}$ with respect to the objective in~(\ref{cemobj}) based on $N_{\mathrm{s}}$ samplings on~(\ref{bernoullisamp}). That is, for each $\boldsymbol{\chi}^{(s)}$, construct corresponding selection matrix $\mathbf{S}_{M_o}^{(s)}$ and compute
\begin{equation}
\label{empd}
\begin{aligned}
    &\mathcal{J}(\mathbf{S}_{M_o}^{(s)}, \boldsymbol{\Phi}) = Q_1 \left( 
    \sqrt{ \frac{2K |\bar{\mathbf{h}}_u^* \mathbf{A}^{(s)} \bar{\mathbf{h}}_f|^2 }{ \|\mathbf{A}^{(s)} \bar{\mathbf{h}}_f \|_2^2 } }, 
    \sqrt{ \frac{ 2(K+1)(2^R-1) }{ \gamma L_f L_u \|\mathbf{A}^{(s)} \bar{\mathbf{h}}_f \|_2^2 } }
    \right),
    \end{aligned}
\end{equation}
where $\mathbf{A}^{(s)} = \mathbf{J}^{\frac{1}{2}} \mathrm{diag}(\boldsymbol{\chi}^{(s)}) \boldsymbol{\Phi} \mathbf{J}^{\frac{1}{2}}$ with the port-indexed $\boldsymbol{\Phi}$ in~(\ref{defa}), so that each port retains its own phase shift regardless of its position in $\mathbf{S}_{M_o}^{(s)}$, and identify $\mathcal{E}^{(t)}$ as the top-$\kappa\%$ of $N_{\mathrm{s}}$ samples (i.e., top-$\lfloor \frac{\kappa N_{\mathrm{s}}}{100} \rfloor$ samples) in $\{\boldsymbol{\chi}^{(s)}\}_{s=1}^{N_{\mathrm{s}}}$ with respect to each $\mathcal{J}(\mathbf{S}_{M_o}^{(s)}, \boldsymbol{\Phi})$ in~(\ref{empd}). The core objective of the CEM is to iteratively update the sampling distribution $\mathbb{P}_{\mathbf{g}^{(t+1)}}$ such that it resembles a distribution concentrated on high-performing solutions. To achieve this, we construct an empirical distribution \(\mathbb{Q}^{(t)}\) from \(\mathcal{E}^{(t)}\):
\begin{equation}
\label{eqd}
\mathbb{Q}^{(t)}(\boldsymbol{\chi}) = \frac{1}{|\mathcal{E}^{(t)}|}  \mathbf{1}_{\{ \boldsymbol{\chi} \in \mathcal{E}^{(t)} \}},
\end{equation}
where \(\mathbf{1}_{\{\cdot\}}\) denotes the indicator function. Then, we minimize the Kullback-Leibler (KL) Divergence \(D_{\mathrm{KL}}(\mathbb{Q}^{(t)} \| \mathbb{P}_{\mathbf{g}^{(t+1)}})\), which effectively encourages the updated \(\mathbb{P}_{\mathbf{g}^{(t+1)}}\) to concentrate more on regions of the solution space that contain promising candidates identified by \(\mathbb{Q}^{(t)}\). Since $D_{\mathrm{KL}}(\mathbb{Q}^{(t)} \| \mathbb{P}_{\mathbf{g}^{(t+1)}})$ differs from the cross-entropy $H(\mathbb{Q}^{(t)},\mathbb{P}_{\mathbf{g}^{(t+1)}})\triangleq- \sum_{\boldsymbol{\chi}} \mathbb{Q}^{(t)}(\boldsymbol{\chi}) \log \mathbb{P}_{\mathbf{g}^{(t+1)}}(\boldsymbol{\chi})$ only by the entropy of $\mathbb{Q}^{(t)}$, which does not depend on $\mathbf{g}^{(t+1)}$, the CEM equivalently updates \(\mathbb{P}_{\mathbf{g}^{(t+1)}}\) by minimizing the cross-entropy:
\begin{equation}
\label{cemopt}
\begin{aligned}
\min_{\mathbf{g}^{(t+1)} \in [0,1]^M} ~& H(\mathbb{Q}^{(t)},\mathbb{P}_{\mathbf{g}^{(t+1)}})\\
 &= - \frac{1}{|\mathcal{E}^{(t)}|} \sum_{\boldsymbol{\chi} \in \mathcal{E}^{(t)}}\sum_{m=1}^M \Biggl(\chi_m \log g_m^{(t+1)} \\
 &~~~~+ (1 - \chi_m) \log (1 - g_m^{(t+1)})\Biggr).
\end{aligned}
\end{equation}
Since this loss function is separable and convex in each \(g_m\), we compute its partial derivative and set it to zero:
\begin{equation}
\label{dklder}
\frac{\partial H}{\partial g_m^{(t+1)}} = 
- \frac{1}{|\mathcal{E}^{(t)}|} \sum_{\boldsymbol{\chi} \in \mathcal{E}^{(t)}} \left(\frac{\chi_m}{g_m^{(t+1)}} - \frac{1 - \chi_m}{1 - g_m^{(t+1)}}\right) = 0.
\end{equation}
Solving this equation yields the optimal update rule for each \(g_m^{(t+1)}\), which is the empirical mean of the elite samples:
\begin{equation}
\tilde{g}_m^{(t+1)} = \frac{1}{|\mathcal{E}^{(t)}|} \sum_{\boldsymbol{\chi} \in \mathcal{E}^{(t)}} \chi_m.
\end{equation}
Then, the selection distribution is updated via exponential moving average by ${g}_m^{(t+1)} \leftarrow \omega \tilde{g}_m^{(t+1)} + (1 - \omega) g_m^{(t)}$~\cite{FRISonoff, CEM}, where $\omega \in (0,1)$ is the smoothing parameter controlling the update rate. 
We update iteratively $\mathbf{g}^{(t)}$ and sample the new feasible solutions $\{\boldsymbol{\chi}^{(s)}\}_{s=1}^{N_{\mathrm{s}}}$ until $\|\mathbf{g}^{(t+1)}-\mathbf{g}^{(t)}\|_{\infty} \le\epsilon_c$ for $\epsilon_c\ll1$. 
After convergence, rather than drawing a fresh sample from $\mathbb{P}_{\mathbf{g}^{(t)}}$, we take the best sample encountered during the whole procedure as the final port selection vector:
\begin{equation}
\label{cembest}
\boldsymbol{\chi}^{\star}=\argmax_{\boldsymbol{\chi}^{(s)}\in\mathcal{C}}~\mathcal{J}(\mathbf{S}_{M_o}^{(s)},\boldsymbol{\Phi}),
\end{equation}
where $\mathcal{C}$ denotes the set of all samples evaluated over the CEM iterations. Since $\mathcal{J}$ is evaluated for every sample in~(\ref{empd}), tracking the best one requires no additional computation. Thereafter, the selection matrix $\mathbf{S}_{M_o}^{\star\mathrm{T}}=[\mathbf{e}_{i_1}\cdots \mathbf{e}_{i_{M_o}}]$ is constructed accordingly by selecting indices $\{i_n\}_{n=1}^{M_o}$ with $\chi_{i_n}^{\star}=1$, and the final distribution $\mathbf{g}^{\star}=\mathbf{g}^{(t)}$ is passed to the next AO iteration. The complete procedure for the proposed CEM for FRIS port selection is summarized in Algorithm~\ref{CEM}.
\begin{algorithm}[t]
\caption{CEM-Based FRIS Port Selection}
\label{CEM}
\begin{algorithmic}[1]
\State \multiline{\textbf{Input:} Port-indexed reflection matrix $\boldsymbol{\Phi}$, sample size $N_{\mathrm{s}}$, stopping criterion $\epsilon_c>0$, elite ratio $\kappa\in(0, 100]$, smoothing parameter $\omega\in(0, 1)$, initialized $\mathbf{g}^{\star}$.} 
\State \textbf{Initialize:} $t \leftarrow 0,~\mathbf{g}^{(0)}\leftarrow{\mathbf{g}^{\star}},~\mathcal{C}\leftarrow\emptyset$
\Repeat
    \State \multiline{Generate $N_{\mathrm{s}}$ binary vectors 
    $\{ \boldsymbol{\chi}^{(s)} \}_{s=1}^{N_{\mathrm{s}}}$ with
    \begin{equation}
    \label{sbv}
    \mathbb{P}_{\mathbf{g}^{(t)}}(\boldsymbol{\chi}) = 
    \prod_{m=1}^M \left( g_m^{(t)\chi_m} (1-g_m^{(t)})^{1-\chi_m} \right).
    \end{equation}}
    \State \multiline{For each $\boldsymbol{\chi}^{(s)}$, enforce $\sum_{m=1}^{M} \chi_m^{(s)} = M_o$ by $\{g_m^{(t)}\}$, and 
 evaluate $\mathbf{S}_{M_o}^{(s)}$ and $\mathcal{J}(\mathbf{S}_{M_o}^{(s)}, \boldsymbol{\Phi})$, and add $\boldsymbol{\chi}^{(s)}$ to $\mathcal{C}$}
    \State \multiline{Identify $\mathcal{E}^{(t)}$ as the top-$\kappa\%$ of $\{\boldsymbol{\chi}^{(s)}\}_{s=1}^{N_{\mathrm{s}}}$ with respect to $\mathcal{J}(\mathbf{S}_{M_o}^{(s)}, \boldsymbol{\Phi})$,
    compute $\tilde{g}_m^{(t+1)}= \frac{1}{|\mathcal{E}^{(t)}|} \sum_{\boldsymbol{\chi} \in \mathcal{E}^{(t)}} \chi_m$, and update via $g_m^{(t+1)} \leftarrow \omega \tilde{g}_m^{(t+1)} + (1-\omega) g_m^{(t)}$.}
    \State $t \leftarrow t+1$.
\Until{$\|\mathbf{g}^{(t)}-\mathbf{g}^{(t-1)}\|_{\infty} \le\epsilon_c$}
\State Select $\boldsymbol{\chi}^{\star}$ by~(\ref{cembest}), construct $\mathbf{S}_{M_o}^{\star}$ accordingly, and set $\mathbf{g}^{\star}\leftarrow\mathbf{g}^{(t)}$ 
\State \Return $\mathbf{S}_{M_o}^{\star}$, $\mathbf{g}^{\star}$
\end{algorithmic}
\end{algorithm}
\subsection{Optimizing $\mathbf{S}_{M_o}$: 2. SDR-Based FRIS Port Selection}
\label{sec:gradient_relaxation}
Given $\boldsymbol{\Phi}$, we aim to optimize $\mathbf{S}_{M_o}$ by proposing an alternative FRIS port selection method based on a continuous relaxation. In other words, instead of the CEM-based sampling method with discrete $\boldsymbol{\chi}\in\{0, 1\}^M$, we optimize a relaxed weight vector $\mathbf{w}  \in [0,1]^M$ that controls the contribution of each FRIS element, and select maximum-$M_o$ indices in optimized $\mathbf{w}^{\star}$. The composite $\mathbf{A}(\mathbf{w})$ is defined as
\begin{equation}
\label{cfo}
\mathbf{A}(\mathbf{w}) \triangleq \mathbf{J}^{\frac{1}{2}} \mathrm{diag}(\mathbf{w}) \boldsymbol{\Phi} \mathrm{diag}(\mathbf{w}) \mathbf{J}^{\frac{1}{2}},
\end{equation}
where $\boldsymbol{\Phi}\in\mathbb{C}^{M\times M}$ is the port-indexed reflection matrix in~(\ref{defa}). By the unit-gain convention for inactive ports, the active ports of the current port selection $\mathbf{S}_{M_o}^{\circ}$ of the AO iteration (with activation vector $\boldsymbol{\chi}^{\circ}$ and parameters $(\lambda_{\circ},\sigma_{\circ}^2)$ in~(\ref{lamsig})), referred to as the incumbent, carry their optimized phase shifts, whereas the remaining ports carry unit gain, i.e.,
\begin{equation}
\label{exphil}
\boldsymbol{\Phi} \leftarrow\mathbf{S}_{M_o}^{\circ\mathrm{T}} \big(\mathbf{S}_{M_o}^{\circ}\boldsymbol{\Phi}\mathbf{S}_{M_o}^{\circ\mathrm{T}}\big) \mathbf{S}_{M_o}^{\circ}-\mathbf{S}_{M_o}^{\circ\mathrm{T}} \mathbf{S}_{M_o}^{\circ}+ \mathbf{I}_M ,
\end{equation}
and a port is effectively removed when its weight $w_m$ is close to 0. Note that for a binary $\mathbf{w}=\boldsymbol{\chi}\in\{0,1\}^M$ with $\mathbf{1}^{\mathrm{T}}\boldsymbol{\chi}=M_o$, $\mathbf{A}(\boldsymbol{\chi})=\mathbf{J}^{\frac{1}{2}}\mathrm{diag}(\boldsymbol{\chi})\boldsymbol{\Phi}\mathbf{J}^{\frac{1}{2}}$ coincides with~(\ref{defa}); hence, the relaxation is exact at every feasible port selection, including the incumbent $\boldsymbol{\chi}^{\circ}$. In the first AO iteration, the initial selection is obtained from the maximum-$M_o$ entries of $\mathbf{w}^{\star}\sim \mathcal{U}[0, 1]^M$.

Under these assumptions, we can reformulate Problem~(\ref{cemobj}) with respect to $\mathbf{w}\in[0, 1]^M$ by leveraging the monotonicity of $Q_1 (a, b)$ in objective~\cite{marQ}, while retaining the port budget through $\mathbf{1}^{\mathrm{T}}(\mathbf{w}\odot\mathbf{w})=M_o$, which reduces to $\mathbf{1}^{\mathrm{T}}\boldsymbol{\chi}=M_o$ for binary $\mathbf{w}$:
\begin{equation}
\label{relaxpre}
\max_{\mathbf{w}\in[0, 1]^M}~\left\{
\frac{|\bar{\mathbf{h}}_u^* \mathbf{A}(\mathbf{w}) \bar{\mathbf{h}}_f|^2}{\| \mathbf{A}(\mathbf{w}) \bar{\mathbf{h}}_f \|_2^2},~ {\| \mathbf{A}(\mathbf{w}) \bar{\mathbf{h}}_f \|_2^2}\right\}~\text{s.t.}~\mathbf{1}^{\mathrm{T}}(\mathbf{w}\odot\mathbf{w})=M_o.
\end{equation}
To handle the inherent multi-objective nature of the problem, we reformulate it using the $\epsilon$-constraint method in Problem~(\ref{relaxpre}), whose solutions are (weakly) Pareto-optimal for Problem~(\ref{relaxpre})~\cite{multio}. Hence, we can get
\begin{equation}
\label{relaxopt}
\begin{aligned}
\max_{\mathbf{w}} ~& f(\mathbf{w}) \triangleq
\frac{\left| \bar{\mathbf{h}}_u^* \mathbf{A}(\mathbf{w}) \bar{\mathbf{h}}_f \right|^2}
{\| \mathbf{A}(\mathbf{w}) \bar{\mathbf{h}}_f \|_2^2}\\
\text{s.t.} ~ 
& \| \mathbf{A}(\mathbf{w}) \bar{\mathbf{h}}_f \|_2^2 \ge M_1,~\mathbf{1}^{\mathrm{T}}(\mathbf{w}\odot\mathbf{w})=M_o,~\mathbf{w}\in[0, 1]^M,
\end{aligned}
\end{equation}
where we treated $M_1^{-1}$ as a role of ``$\epsilon$'' in the $\epsilon$-constraint method. Rather than fixing $M_1$ a priori, we set it to the power of the incumbent, i.e., $M_1\triangleq\|\mathbf{A}(\boldsymbol{\chi}^{\circ})\bar{\mathbf{h}}_f\|_2^2=(K+1)\sigma_{\circ}^2$. Then, $\boldsymbol{\chi}^{\circ}$ is feasible for~\eqref{relaxopt}, and any optimal solution $\mathbf{w}$ of~\eqref{relaxopt} satisfies $f(\mathbf{w})\ge f(\boldsymbol{\chi}^{\circ})=\lambda_{\circ}/K$ and $\|\mathbf{A}(\mathbf{w})\bar{\mathbf{h}}_f\|_2^2\ge(K+1)\sigma_{\circ}^2$, i.e., the continuous counterparts of $\lambda$ and $\sigma^2$ in~(\ref{lamsig}) do not decrease relative to the incumbent. Problem~\eqref{relaxopt} is still nonconvex since it involves a fractional quadratic objective. To handle this difficulty, we adopt the SDR followed by Gaussian randomization~\cite{sdr, FRISlook}. Starting from~\eqref{relaxopt}, define $\mathbf x \triangleq \mathbf w \odot \mathbf w \in \mathbb R^M$ with Hadamard Product $\odot$, so that both numerator and denominator of $f$ can be expressed as:
\begin{equation}
\label{numden}
\begin{aligned}
&\big|\bar{\mathbf h}_u^{*}\mathbf A(\mathbf w)\bar{\mathbf h}_f\big|^2
= \mathbf x^{\mathrm{T}} \mathbf C \mathbf x=\mathrm{tr}(\mathbf C\mathbf X)\\
&\|\mathbf A(\mathbf w)\bar{\mathbf h}_f\|_2^2
= \mathbf x^{\mathrm{T}} \mathbf D \mathbf x=\mathrm{tr}(\mathbf D\mathbf X), 
\end{aligned}
\end{equation}
where $\mathbf X=\mathbf x\mathbf x^{\mathrm{T}}\succeq 0$ with $\mathrm{rank}(\mathbf X)=1$ and
\begin{equation}
\label{defcc}
\begin{aligned}
&\mathbf C \triangleq \mathbf c \mathbf c^*,~\mathbf c \triangleq \mathrm{diag}(\boldsymbol{\Phi})\odot(\mathbf J^{\frac{1}{2}}\bar{\mathbf h}_f)\odot\overline{(\mathbf J^{\frac{1}{2}}\bar{\mathbf h}_u)}\\
&\mathbf D \triangleq \mathrm{diag}\big(\mathbf J^{\frac{1}{2}}\bar{\mathbf h}_f \odot \mathrm{diag}(\boldsymbol{\Phi})\big)^*\mathbf J \mathrm{diag}\big(\mathbf J^{\frac{1}{2}}\bar{\mathbf h}_f \odot \mathrm{diag}(\boldsymbol{\Phi})\big).
\end{aligned}
\end{equation}
Note that $\mathbf C, \mathbf D\succeq 0$. Moreover, the port budget $\mathbf{1}^{\mathrm{T}}\mathbf{x}=M_o$ is lifted as $\mathbf{1}^{\mathrm{T}}\mathbf{X}\mathbf{1}=(\mathbf{1}^{\mathrm{T}}\mathbf{x})^2=M_o^2$, and since $0\le x_m\le 1$ implies $x_m^2\le x_m$, we also have the valid inequality $\mathrm{tr}(\mathbf{X})=\|\mathbf{x}\|_2^2\le\mathbf{1}^{\mathrm{T}}\mathbf{x}=M_o$, which holds with equality for binary $\mathbf{x}$ and tightens the subsequent relaxation. Thus, Problem~\eqref{relaxopt} can be equivalently reformulated as
\begin{equation}
\label{eq:rank1_form}
\begin{aligned}
\max_{\mathbf X\succeq 0}~&
 \frac{\mathrm{tr}(\mathbf C\mathbf X)}{\mathrm{tr}(\mathbf D\mathbf X)}~\text{s.t.}~\mathrm{tr}(\mathbf D\mathbf X) \ge M_1,~\mathrm{rank}(\mathbf X)=1,\\
 &\mathbf{1}^{\mathrm{T}}\mathbf{X}\mathbf{1}=M_o^2,~\mathrm{tr}(\mathbf{X})\le M_o,~\forall X_{ij}\in [0,1].
\end{aligned}
\end{equation}
The rank-one constraint ($\mathrm{rank}(\mathbf X)=1$) renders \eqref{eq:rank1_form} non-convex. Hence, we relax this constraint and employ SDR. Furthermore, to eliminate the fractional objective in~\eqref{eq:rank1_form}, we apply the Charnes-Cooper transformation by $\mathbf Y \triangleq \frac{\mathbf X}{\mathrm{tr}(\mathbf D\mathbf X)}$ and $s \triangleq \frac{1}{\mathrm{tr}(\mathbf D\mathbf X)}$. This yields the following linear semidefinite programming (SDP)~\cite{charnes}:
\begin{equation}
\label{eq:CC_SDR_link}
\begin{aligned}
\max_{\mathbf Y\succeq 0,~s}~ 
& \mathrm{tr}(\mathbf C\mathbf Y)\\
\text{s.t.}~ & \mathrm{tr}(\mathbf D\mathbf Y)=1,~\forall Y_{ij}\in[0, s],~0<s \le M_1^{-1},\\
& \mathbf{1}^{\mathrm{T}}\mathbf{Y}\mathbf{1}=M_o^2 s,~\mathrm{tr}(\mathbf{Y})\le M_o s,
\end{aligned}
\end{equation}
which provides an efficiently computable upper-bound on the optimal value of the rank-constrained~\eqref{eq:rank1_form}~\cite{boyd}. Since the lifted incumbent $\boldsymbol{\chi}^{\circ}\boldsymbol{\chi}^{\circ\mathrm{T}}$ is feasible for~\eqref{eq:rank1_form} under $M_1=(K+1)\sigma_{\circ}^2$, Problem~\eqref{eq:CC_SDR_link} is always feasible, and its optimal value is no smaller than $f(\boldsymbol{\chi}^{\circ})$.
A feasible $\mathbf w$ can then be reconstructed from the solution $(\mathbf Y^\star,~s^\star)$ by forming $\mathbf X^\star=\frac{\mathbf Y^\star}{s^\star}$, factorizing it as $\mathbf x\mathbf x^{\mathrm{T}}$, and finally obtaining $\mathbf w$ via the element-wise square root, i.e., $\mathbf w=\sqrt{\mathbf x}$. In general, however, $\mathbf X^\star$ does not satisfy the rank-one condition, and thus cannot be factorized directly as $\mathbf x\mathbf x^{\mathrm{T}}$. Hence, we adopt the Gaussian randomization to recover $\mathbf w$~\cite{sdr}. We interpret $\mathbf X^\star$ as a covariance and draw $N_{\mathrm{r}}$ independent and identically distributed (i.i.d.) samples:
\begin{equation}
\label{iidz}
\mathbf r_\ell \sim \mathcal{N}(\mathbf 0,\mathbf X^\star)~(\ell=1,\cdots,N_{\mathrm{r}}),
\end{equation}
which is done by the eigendecomposition $\mathbf X^\star=\mathbf U\boldsymbol{\Lambda}\mathbf U^{\mathrm{T}}$ and letting $\mathbf r_\ell = \mathbf U\boldsymbol{\Lambda}^{\frac{1}{2}} \bar{\mathbf r}_\ell$ with $\bar{\mathbf r}_\ell\sim \mathcal{N} (0, \mathbf I_M)$~\cite{chol}; unlike the Cholesky decomposition, this remains valid when $\mathbf X^\star$ is rank-deficient, which is typical for SDR solutions. Thereafter, since $\mathbf X^\star$ cannot distinguish $\mathbf x$ from
$-\mathbf x$, i.e., $\mathbf x\mathbf x^{\mathrm T}=(-\mathbf x)(-\mathbf x)^{\mathrm T}$,
whereas $\mathbf x\ge\mathbf 0$ with $\mathbf 1^{\mathrm T}\mathbf x=M_o>0$,
each $\mathbf r_\ell$ is first sign-aligned as
$\hat{\mathbf r}_\ell\triangleq\mathrm{sgn}(\mathbf 1^{\mathrm T}\mathbf r_\ell)\,\mathbf r_\ell$
and then projected elementwise onto $[0,1]^M$ to form a candidate:
\begin{equation}
\label{projw}
\tilde{\mathbf{x}}_\ell \leftarrow \arg\min_{\mathbf x} \|\mathbf x - \hat{\mathbf r}_\ell\|_2^2~(\mathbf{x}\in[0, 1]^M),~\tilde{\mathbf{w}}_\ell =\sqrt{\tilde{\mathbf{x}}_\ell}.
\end{equation}
For each repaired $\tilde{\mathbf w}_\ell$, we directly form the binary candidate $\tilde{\boldsymbol{\chi}}_\ell$ by allocating 1 to the maximum-$M_o$ entries of $\tilde{\mathbf w}_\ell$ and 0 otherwise, so that the candidates are assessed on the actual port configurations rather than on their continuous surrogates. Let $\tilde{\mathbf{S}}_{M_o,\ell}$ denote the selection matrix of $\tilde{\boldsymbol{\chi}}_\ell$, and define the set of candidates satisfying the $\epsilon$-constraint at the binary level as $\mathcal{K}\triangleq\{\tilde{\boldsymbol{\chi}}_\ell:\sigma^2(\tilde{\mathbf{S}}_{M_o,\ell},\boldsymbol{\Phi})\ge\sigma_{\circ}^2\}$. The final selection is
\begin{equation}
\label{sdrbest}
\boldsymbol{\chi}^{\star}=\begin{cases}\argmax_{\boldsymbol{\chi}\in\mathcal{K}}~\lambda(\mathbf{S}_{M_o}(\boldsymbol{\chi}),\boldsymbol{\Phi}), & \mathcal{K}\neq\emptyset,\\ \argmax_{\ell}~\sigma^2(\tilde{\mathbf{S}}_{M_o,\ell},\boldsymbol{\Phi}), & \mathcal{K}=\emptyset,\end{cases}
\end{equation}
where $\mathbf{S}_{M_o}(\boldsymbol{\chi})$ is the selection matrix of $\boldsymbol{\chi}$; i.e., the objective of~\eqref{relaxopt} is maximized among the candidates satisfying its $\epsilon$-constraint, and the candidate closest to feasibility is selected otherwise. The resulting selection matrix is $\mathbf{S}_{M_o}^{\star\mathrm{T}}=[\mathbf{e}_{i_1} \cdots \mathbf{e}_{i_{M_o}}]$ with $\chi_{i_n}^{\star}=1$. The complete procedure is summarized in Algorithm~\ref{relaxalg}.
\begin{algorithm}[t]
\caption{SDR-Based FRIS Port Selection}
\label{relaxalg}
\begin{algorithmic}[1]
\State \multiline{\textbf{Input:} Port-indexed reflection matrix $\boldsymbol{\Phi}$, incumbent $\mathbf{S}_{M_o}^{\circ}$ ($\boldsymbol{\chi}^{\circ}$) with $\sigma_{\circ}^2$, sample size $N_{\mathrm{r}}$}
\State Set $M_1\leftarrow\|\mathbf{A}(\boldsymbol{\chi}^{\circ})\bar{\mathbf{h}}_f\|_2^2$ and $\mathcal{K}\leftarrow\emptyset$
\State Solve the relaxed SDP~\eqref{eq:CC_SDR_link} to obtain $\mathbf X^\star\succeq 0$
\For{$\ell=1,\cdots,N_{\mathrm{r}}$}
    \State Draw random vector $\mathbf r_\ell \sim \mathcal{N}(\mathbf 0,\mathbf X^\star)$
    \State $\tilde{\mathbf{x}}_\ell \leftarrow \arg\min_{\mathbf x} \|\mathbf x - \mathbf r_\ell\|_2^2~(\mathbf{x}\in[0, 1]^M),~\tilde{\mathbf{w}}_\ell =\sqrt{\tilde{\mathbf{x}}_\ell}$
    \State \multiline{Round $\tilde{\mathbf{w}}_\ell$ to $\tilde{\boldsymbol{\chi}}_\ell$ (maximum-$M_o$ entries $\to1$), evaluate~(\ref{lamsig}), and add $\tilde{\boldsymbol{\chi}}_\ell$ to $\mathcal{K}$ if $\sigma^2(\tilde{\mathbf{S}}_{M_o,\ell},\boldsymbol{\Phi})\ge\sigma_{\circ}^2$}
\EndFor
\State Select $\boldsymbol{\chi}^{\star}$ by~\eqref{sdrbest}
\State \multiline{$\mathbf{S}_{M_o}^{\star\mathrm{T}}=[\mathbf{e}_{i_1} \cdots \mathbf{e}_{i_{M_o}}]$: Select the $\{i_n\}_{n=1}^{M_o}$th columns of $\mathbf{I}_M$ with $\chi_{i_n}^{\star}=1$.}
\State \Return $\mathbf{S}_{M_o}^{\star}$
\end{algorithmic}
\end{algorithm}
\subsection{Optimizing $\boldsymbol{\Phi}$: Dinkelbach-Based Riemannian Optimization with Penalty}\label{phi}
Similar to~(\ref{relaxpre}), for fixed $\mathbf{S}_{M_o}$, we reformulate Problem~(\ref{objoriginal}) with respect to $\boldsymbol{\Phi}$ by leveraging the monotonicity of $Q_1 (a, b)$ in objective~\cite{marQ}:
\begin{equation}
\label{multiobj}
\begin{aligned}
\max_{\boldsymbol{\Phi}}&~\left\{
\frac{|\bar{\mathbf{h}}_u^* \mathbf{A} \bar{\mathbf{h}}_f|^2}{\| \mathbf{A} \bar{\mathbf{h}}_f \|_2^2},~ {\| \mathbf{A} \bar{\mathbf{h}}_f \|_2^2}
\right\}\\
\text{s.t.} ~ & 
\mathbf{S}_{M_o} \in \mathcal{P}_{M_o},~ \boldsymbol{\Phi} \triangleq \mathrm{diag}(\{e^{j\phi_m}\}_{m=1}^{M})~(\forall \phi_m \in [0, 2\pi)).
\end{aligned}
\end{equation}
Again, by applying the $\epsilon$-constraint method~\cite{multio} for multiple objectives in~(\ref{multiobj}), we can get
\begin{equation}
\label{phaseopt}
\begin{aligned}
\max_{\boldsymbol{\Phi}}  &~ \frac{|\bar{\mathbf{h}}_u^* \mathbf{A} \bar{\mathbf{h}}_f|^2}{\|\mathbf{A} \bar{\mathbf{h}}_f\|_2^2} \\
\text{s.t.} ~
&{\|\mathbf{A} \bar{\mathbf{h}}_f\|_2^2} \ge M_2,~\mathbf{S}_{M_o} \in \mathcal{P}_{M_o},\\
&\boldsymbol{\Phi} \triangleq \mathrm{diag}(\{e^{j\phi_m}\}_{m=1}^{M})~(\forall\phi_m \in [0, 2\pi)),
\end{aligned}
\end{equation}
where same as in~(\ref{relaxopt}), we treat $M_2^{-1}$ as a role of ``$\epsilon$'' in the $\epsilon$-constraint method. Since only the active ports affect $\mathbf{A}$, the optimization variable of~(\ref{phaseopt}) is the unit-modulus vector of the active phase shifts $\mathbf{z} \triangleq \mathbf{S}_{M_o}\mathrm{diag}(\boldsymbol{\Phi}) = [e^{j\phi_{i_1}} \cdots e^{j\phi_{i_{M_o}}}]^{\mathrm{T}}\in \mathbb{C}^{M_o}~(\forall |z_n| = 1)$, which satisfies $\mathrm{diag}(\boldsymbol{\chi})\boldsymbol{\Phi}=\mathbf{S}_{M_o}^{\mathrm{T}}\mathrm{diag}(\mathbf{z})\mathbf{S}_{M_o}$; the inactive entries of $\boldsymbol{\Phi}$ are kept at unit gain. As in~(\ref{relaxopt}), we set $M_2$ to the power of the incumbent, i.e., $M_2\triangleq\|\mathbf{A}(\mathbf{S}_{M_o},\boldsymbol{\Phi}^{\circ})\bar{\mathbf{h}}_f\|_2^2$, where $\boldsymbol{\Phi}^{\circ}$ is the incumbent reflection matrix, so that the incumbent active-phase vector $\mathbf{z}^{\circ}\triangleq\mathbf{S}_{M_o}\mathrm{diag}(\boldsymbol{\Phi}^{\circ})$ is feasible for~(\ref{phaseopt}) and serves as the warm start of the phase optimization. Let us define $\mathbf{b}_u \triangleq \mathbf{S}_{M_o} \mathbf{J}^{\frac{1}{2}} \bar{\mathbf{h}}_u$ and $\mathbf{b}_f \triangleq \mathbf{S}_{M_o} \mathbf{J}^{\frac{1}{2}} \bar{\mathbf{h}}_f$. Then, the numerator of the objective function in~(\ref{phaseopt}) can be equivalently:
\begin{equation}
    |\bar{\mathbf{h}}_u^* \mathbf{A} \bar{\mathbf{h}}_f |^2
    = |\mathbf{b}_u^* \operatorname{diag}(\mathbf{z}) \mathbf{b}_f|^2
    = |\mathbf{b}_u^* \operatorname{diag}(\mathbf{b}_f) \mathbf{z}|^2.
\end{equation}
Similarly, the denominator becomes
\begin{equation}
    \|\mathbf{A} \bar{\mathbf{h}}_f\|_2^2 = 
    \left\| \mathbf{J}^{\frac{1}{2}} \mathbf{S}_{M_o}^{\mathrm{T}} \operatorname{diag}(\mathbf{z}) \mathbf{b}_f \right\|_2^2
    = \mathbf{z}^* \mathbf{Q} \mathbf{z},
\end{equation}
where $\mathbf{Q} \triangleq \operatorname{diag}(\mathbf{b}_f^*) \mathbf{S}_{M_o} \mathbf{J} \mathbf{S}_{M_o}^{\mathrm{T}}\operatorname{diag}(\mathbf{b}_f)$. 
Therefore, the original constrained optimization problem becomes:
\begin{equation}
\label{riemannianpower}
\begin{aligned}
&\max_{\mathbf{z}}~\frac{|\mathbf{a}^* \mathbf{z}|^2}{\mathbf{z}^* \mathbf{Q}\mathbf{z}}~\text{s.t.}~\mathbf{z}^*\mathbf{Q}\mathbf{z}\ge M_2,~\mathbf{z}\in\mathbb{T}^{M_o},
\end{aligned}
\end{equation}
where $\mathbf{a}=\mathrm{diag}(\mathbf{b}_f^*)\mathbf{b}_u$ and $\mathbb{T}^{M_o} \triangleq \left\{ \mathbf{z} \in \mathbb{C}^{M_o} : |z_i| = 1~(\forall i) \right\}$ denotes the complex unit-modulus torus. To address the fractional quadratic objective, we employ a Dinkelbach-based Riemannian gradient-ascent framework~\cite{Din, Rieman1}. To start, let $N(\mathbf z)\triangleq|\mathbf a^*\mathbf z|^2=\mathbf z^*\mathbf R_a \mathbf z~(\mathbf R_a\triangleq\mathbf a\mathbf a^*)$ and $D(\mathbf z)\triangleq\mathbf z^*\mathbf Q\mathbf z>0$. Thereafter, Problem~\eqref{riemannianpower} can be tackled by the Dinkelbach transform on $\mathbb T^{M_o}$~\cite{Din}:
\begin{equation}
\label{eq:dink_subprob}
\max_{\mathbf z}~N(\mathbf z)-\tau D(\mathbf z)~\text{s.t.}~ D(\mathbf z)\ge M_2,~\mathbf z\in\mathbb T^{M_o},
\end{equation}
with the outer update $\tau_{k+1}= \frac{N(\mathbf z_{k+1})}{D(\mathbf z_{k+1})}$, where $\mathbf z_{k+1}$ denotes the solution of~\eqref{eq:dink_subprob} obtained for $\tau=\tau_k$.  We enforce $D(\mathbf z)\ge M_2$ via a quadratic (squared-hinge) penalty:
\begin{equation}
\label{eq:penalty_obj}
F_{\tau,\mu_0}(\mathbf z) \triangleq N(\mathbf z)-\tau D(\mathbf z)-\mu_0[M_2-D(\mathbf z)]_+^2,
\end{equation}
where $[t]_+\triangleq\max\{t,0\}$ and the penalty parameter $\mu_0>0$ grows when the constraint is violated. Unlike the non-smooth penalty $[M_2-D(\mathbf z)]_+$, the squared-hinge penalty is continuously differentiable, so that the Riemannian gradient below is well defined everywhere and Armijo backtracking applies. Hence, the penalized problem with Dinkelbach transform is: 
\begin{equation}
\label{eq:riem_main}
\max_{\mathbf z \in\mathbb T^{M_o}} ~F_{\tau,\mu_0}(\mathbf z).
\end{equation}
Since the feasible set is the unit-modulus complex torus, we employ Riemannian gradient ascent to solve~\eqref{eq:riem_main}. 
First, $\nabla F_{\tau,\mu_0}$ can be formulated as
\begin{equation}
\label{eug22}
\nabla F_{\tau,\mu_0}(\mathbf z)=2(\mathbf R_a-\tau\mathbf Q)\mathbf z+4\mu_0[M_2-D(\mathbf z)]_+\mathbf Q\mathbf z.
\end{equation}
Projecting onto the tangent space $\mathcal T_{\mathbf z}\mathbb T^{M_o}\triangleq\{\boldsymbol\xi: \Re\{\boldsymbol\xi\odot\overline{\mathbf z}\}=0\}$ with respect to $\mathbf{z}$ gives the Riemannian gradient~\cite{Rieman1, Rieman2}
\begin{equation}
\label{remg}
\mathrm{grad}_{\mathbb T^{M_o}} F_{\tau,\mu_0}(\mathbf z)=\nabla F_{\tau,\mu_0}(\mathbf z)-\Re\{\nabla F_{\tau,\mu_0}(\mathbf z)\odot\overline{\mathbf z}\}\odot \mathbf z,
\end{equation}
and we use the retraction mapping $\mathcal R_{\mathbf z}(\boldsymbol\xi): \mathcal{T}_{\mathbf{z}} \mathbb{T}^{M_o} \rightarrow  \mathbb{T}^{M_o}$ to restrict $\boldsymbol{\xi}\in\mathbb{T}^{M_o}$, which is a smooth mapping that satisfies the following conditions~\cite{Rieman1}:
\begin{enumerate}
\item $R_{\mathbf{z}} (\mathbf{0_z})=\mathbf{z}$; $\mathbf{0_z}\in\mathcal{T}_{\mathbf{z}} \mathbb{T}^{M_o}$ is the zero-element of $\mathcal{T}_{\mathbf{z}} \mathbb{T}^{M_o} $
\item $\mathrm{D}R_{\mathbf{z}} (\mathbf{0_z})=\mathrm{Id}_{\mathcal{T}_{\mathbf{z}} \mathbb{T}^{M_o}}$, i.e., $\frac{d}{dt} R_{\mathbf{z}} (t \boldsymbol{\zeta})|_{t=0} = \mathcal{T}_{\mathbf{0_z}}R_{\mathbf{z}} (\boldsymbol{\zeta}) = \boldsymbol{\zeta}$ for all $\boldsymbol{\zeta} \in \mathcal{T}_{\mathbf{z}} \mathbb{T}^{M_o}$ and $t\in\mathbb{R}$, where $\mathrm{D}$ is a tangent map and $\mathrm{Id}_{\mathcal{T}_{\mathbf{z}} \mathbb{T}^{M_o}}$ is the identity mapping on $\mathcal{T}_{\mathbf{z}} \mathbb{T}^{M_o}$
\end{enumerate}
Hence, with an adaptive step size $\eta_k>0$ from Armijo backtracking~\cite{Rieman2}, the ascent update reads
\begin{equation}
\label{eq:riem_update}
\mathbf z_{k+1}= \mathcal R_{\mathbf z_k}\Big(\eta_k\mathrm{grad}_{\mathbb T^{M_o}}F_{\tau,\mu_0}\big(\mathbf z_k\big)\Big),
\end{equation}
until $\|\mathrm{grad}_{\mathbb T^{M_o}}F_{\tau_k,\mu_0}(\mathbf z)\|_2$ shrinks down certain threshold. Since $\mathbb T^{M_o}$ is the unit-modulus complex torus, its manifold geometry allows us to characterize $R_{\mathbf z_k}\Big(\eta_k\mathrm{grad}_{\mathbb T^{M_o}}F_{\tau,\mu_0}\big(\mathbf z_k\big)\Big)$ explicitly by Lemma~\ref{rettheo} and Corollary~\ref{retcor}.

\begin{lemma}
\label{rettheo}
For given $\boldsymbol{\xi}\in\mathcal{T}_{\mathbf{z}} \mathbb{T}^{M_o}$, there exists corresponding $\boldsymbol{\theta}\in\mathbb{R}^{M_o}$ such that $R_{\mathbf{z}} (\boldsymbol{\xi}) =e^{j \boldsymbol{\theta}} \odot \mathbf{z}$, where $\boldsymbol\theta=\Im (\frac{\boldsymbol\xi }{ \mathbf{z}})$ and $\frac{\mathbf{a}}{\mathbf{b}}$ is the element-wise division of $\mathbf{a}$ and $\mathbf{b}$
\end{lemma}
\begin{proof}
For each $i$, the unit-modulus constraint requires $|z_i| = 1$, so $\mathbb{T}^{M_o}$ can be viewed as a product of $M_o$ copies of the complex unit circle. Moreover, by definition of $\mathcal{T}_{\mathbf{z}} \mathbb{T}^{M_o}$, for $\boldsymbol{\xi}\in\mathcal{T}_{\mathbf{z}} \mathbb{T}^{M_o}$, 
$\xi_i$ must be perpendicular to $z_i$ in the complex plane, so any $\boldsymbol{\xi}$ can be written as:
\begin{equation}
\label{xiphase}
\boldsymbol{\xi} = j  \boldsymbol{\theta} \odot \mathbf{z}~(\exists \boldsymbol{\theta} \in \mathbb{R}^{M_o}),
\end{equation}
and by dividing with $\mathbf z$ element-wise, we get  $\boldsymbol\theta=\Im \left(\frac{\boldsymbol\xi }{ \mathbf{z}}\right)$. Since by~(\ref{xiphase}), each $\xi_i$ corresponds to a phase rotation of $z_i$ by $\frac{\pi}{2}$ and multiply it by $\theta_i$, the most natural retraction is:
\begin{equation}
\label{natret}
R_{\mathbf{z}} (\boldsymbol{\xi}) =e^{j \boldsymbol{\theta}} \odot \mathbf{z},
\end{equation}
where $e^{j \boldsymbol{\theta}} \triangleq\left[ e^{j \theta_1} \cdots e^{j \theta_{M_o}} \right]^{\mathrm{T}}$, which clearly satisfies $|[R_{\mathbf{z}} (\boldsymbol{\xi})]_i| = |z_i| = 1$ for all $i$, thus ensuring that $R_{\mathbf{z}} (\boldsymbol{\xi})\in \mathbb{T}^{M_o}$.
Furthermore, by considering $t\boldsymbol{\xi}~(t\in\mathbb{R})$, we can get
\begin{equation}
\label{defcor}
\frac{d}{dt} R_{\mathbf{z}} (t \boldsymbol{\xi})|_{t=0} =\frac{d}{dt}( e^{j t\boldsymbol{\theta}} \odot \mathbf{z}) |_{t=0} = j  \boldsymbol{\theta} \odot \mathbf{z}= \boldsymbol{\xi},
\end{equation}
which ensures $\mathrm{D}R_{\mathbf{z}} (\mathbf{0_z})=\mathrm{Id}_{\mathcal{T}_{\mathbf{z}} \mathbb{T}^{M_o}}$, and the lemma follows.
\end{proof}
\begin{corollary}
\label{retcor}
$R_{\mathbf z_k}\Big(\eta_k\mathrm{grad}_{\mathbb T^{M_o}}F_{\tau,\mu_0}\big(\mathbf z_k\big)\Big)$ is given by:
\begin{equation}
\label{retcorr}
R_{\mathbf z_k}\Big(\eta_k\mathrm{grad}_{\mathbb T^{M_o}}F_{\tau,\mu_0}\big(\mathbf z_k\big)\Big)
= e^{ j  \Im\left\{ \frac{\eta_k\mathrm{grad}_{\mathbb T^{M_o}}F_{\tau,\mu_0}(\mathbf z_k)}{ \mathbf{z}_k } \right\} } \odot \mathbf{z}_k.
\end{equation}
\end{corollary}

The complete steps of the proposed Dinkelbach-based Riemannian algorithm are outlined in Algorithm~\ref{adp}.
\begin{algorithm}[t]
\caption{Dinkelbach-based Riemannian Algorithm with Penalty}
\label{adp}
\begin{algorithmic}[1]
\State \multiline{\textbf{Input:} Selection $\mathbf{S}_{M_o}$, incumbent $\boldsymbol{\Phi}^{\circ}$, tolerance $\epsilon_r, \text{tol}>0$, penalty parameter $\mu_0>0, \alpha_0>1$}
\State \multiline{\textbf{Initialize:} $k\gets 0$, $\mathbf z_0\gets\mathbf{z}^{\circ}=\mathbf{S}_{M_o}\mathrm{diag}(\boldsymbol{\Phi}^{\circ})$, $M_2\gets D(\mathbf{z}_0)$, $\tau_0\gets N(\mathbf{z}_0)/D(\mathbf{z}_0)$}
\While{not converged}
\State \multiline{Starting from $\mathbf z_k$, maximize $F_{\tau_k,\mu_0}(\mathbf z)$ on $\mathbb T^{M_o}$ by $\mathbf z'\gets \mathcal R_{\mathbf z}\Big(\eta\mathrm{grad}_{\mathbb T^{M_o}}F_{\tau,\mu_0}\big(\mathbf z\big)\Big)$, and obtain solution $\widehat{\mathbf z}$}
\If {$D(\widehat{\mathbf z})<M_2-\text{tol}$}
\State $\mu_0 \gets \alpha_0 \mu_0$
\EndIf
\State  $\tau_{k+1} \gets \frac{N(\widehat{\mathbf z})}{D(\widehat{\mathbf z})},~\mathbf z_{k+1} \gets \widehat{\mathbf z},~k \gets k+1$
\If {$|\tau_{k}-\tau_{k-1}|\le \epsilon_r$ and $D(\mathbf z_{k})\ge M_2-\text{tol}$}
\State break
\EndIf
\EndWhile
\State $\mathbf z^\star \gets \mathbf z_k$
\State \Return $\boldsymbol{\Phi}^\star=\mathbf{S}_{M_o}^{\mathrm{T}}\mathrm{diag}(\mathbf{z}^\star)\mathbf{S}_{M_o}+\mathbf{I}_M-\mathbf{S}_{M_o}^{\mathrm{T}}\mathbf{S}_{M_o}$
\end{algorithmic}
\end{algorithm}
\subsection{Proposed AO Frameworks: AO-CEM and AO-SDR}\label{sec:ao_framework}
We now present two independent AO frameworks that iteratively update $\mathbf{S}_{M_o}$ with the CEM-based FRIS port selection (via Algorithm~\ref{CEM}; AO-CEM), or the SDR-based FRIS port selection (via Algorithm~\ref{relaxalg}; AO-SDR) and the Dinkelbach-based Riemannian method with penalty for optimizing $\boldsymbol{\Phi}$ (via Algorithm~\ref{adp}), presented in Algorithm~\ref{aocem} and~\ref{aorelax}, respectively. In the $i$th iteration, the SDR sub-step of AO-SDR is warm-started at the
incumbent $(\mathbf{S}_{M_o}^{(i)},\boldsymbol{\Phi}^{(i)})$ with $M_1$ set
to its power, whereas the phase sub-step of both frameworks is warm-started
at $(\mathbf{S}_{M_o}^{(i+1)},\boldsymbol{\Phi}^{(i)})$, i.e., the updated
port selection with the previous phase shifts, with $M_2$ set to its power;
the resulting $(\mathbf{S}_{M_o}^{(i+1)},\boldsymbol{\Phi}^{(i+1)})$ serves
as the incumbent of the next iteration.
We repeat this process until a predefined stopping criterion $\big|\mathcal{J}^{(i+1)} - \mathcal{J}^{(i)}\big| \le \delta_m~(m=1, 2)$ is met. We will show in Section~\ref{numer} that the proposed AO frameworks exhibit steady improvement and converge within a modest number of iterations in practice.\footnote{Since $M_1$ and $M_2$ are set to the power of the respective incumbents,
each incumbent is feasible for its subproblem, i.e.,~(\ref{relaxopt})
and~(\ref{phaseopt}), respectively. Hence, if each subproblem were solved exactly, no sub-step would decrease $\mathcal{J}$: the CEM maximizes $\mathcal{J}$ directly, while the SDR and Riemannian sub-steps would decrease neither $\lambda$ nor $\sigma^2$, in each of which $\mathcal{J}$ is non-decreasing; the bounded sequence $\{\mathcal{J}^{(i)}\}$ would then converge. In practice, however, the subproblems are solved approximately, i.e., via relaxation, randomization, and rounding in the SDR, stochastic sampling in the CEM, and a penalty-based stationary point in the Riemannian method; thus, strict monotonicity is not guaranteed analytically but verified numerically.} Moreover, since $\mathbf{S}_{M_o}$ is discrete and~(\ref{objoriginal}) is non-convex, global optimality cannot be assured; we demonstrate in Section~\ref{numer} that the proposed AO frameworks consistently yield near-optimal performance compared to the result obtained by BFS.
\begin{algorithm}[t]
\caption{AO-CEM: Alternating Optimization with CEM-Based Port Selection}
\label{aocem}
\begin{algorithmic}[1]
\State \multiline{\textbf{Input:} Initial port-indexed $\boldsymbol{\Phi}^{(0)}=\mathrm{diag}(\{e^{j\phi_m}\}_{m=1}^{M})~(\forall \phi_m \sim\mathcal U [0, 2\pi)),~\mathbf{g}^{\star}\sim \mathcal{U}[0, 1]^M$, stopping criterion $\delta_1$.}
\State \multiline{$i \leftarrow 0$, evaluate $\mathbf{S}_{M_o}^{(0)}$ by $\mathbf{g}^{\star}$, reset the inactive entries of $\boldsymbol{\Phi}^{(0)}$ to unit gain, and compute $\mathcal{J}^{(0)} = \mathcal{J}(\mathbf{S}_{M_o}^{(0)}, \boldsymbol{\Phi}^{(0)})$.}
\While{$\big|\mathcal{J}^{(i+1)} - \mathcal{J}^{(i)}\big| > \delta_1$}
    \State \multiline{Fix $\boldsymbol{\Phi}^{(i)}$ and update $\mathbf{S}_{M_o}^{(i+1)}$ via Algorithm~\ref{CEM}.}
        \State \multiline{Fix $\mathbf{S}_{M_o}^{(i+1)}$ and update $\boldsymbol{\Phi}^{(i+1)}$ via Algorithm~\ref{adp} with incumbent $\boldsymbol{\Phi}^{\circ}=\boldsymbol{\Phi}^{(i)}$.}
    \State Compute $\mathcal{J}^{(i+1)} = \mathcal{J}(\mathbf{S}_{M_o}^{(i+1)}, \boldsymbol{\Phi}^{(i+1)})$.
     \State $i \leftarrow i+1$
\EndWhile
\State \Return $(\boldsymbol{\Phi}^{(i)}, \mathbf{S}_{M_o}^{(i)})$.
\end{algorithmic}
\end{algorithm}
\begin{algorithm}[t]
\caption{AO-SDR: Alternating Optimization with SDR-Based Port Selection}
\label{aorelax}
\begin{algorithmic}[1]
\State \multiline{\textbf{Input:} Initial port-indexed $\boldsymbol{\Phi}^{(0)}=\mathrm{diag}(\{e^{j\phi_m}\}_{m=1}^{M})~(\forall \phi_m \sim\mathcal U [0, 2\pi)),~\mathbf{w}^{\star}\sim \mathcal{U}[0, 1]^M$, stopping criterion $\delta_2$.}
\State \multiline{$i \leftarrow 0$, evaluate $\mathbf{S}_{M_o}^{(0)}$ by $\mathbf{w}^{\star}$, reset the inactive entries of $\boldsymbol{\Phi}^{(0)}$ to unit gain, and compute $\mathcal{J}^{(0)} = \mathcal{J}(\mathbf{S}_{M_o}^{(0)}, \boldsymbol{\Phi}^{(0)})$.}
\While{$\big|\mathcal{J}^{(i+1)} - \mathcal{J}^{(i)}\big| > \delta_2$}
      \State \multiline{Fix $\boldsymbol{\Phi}^{(i)}$ and update $\mathbf{S}_{M_o}^{(i+1)}$ via Algorithm~\ref{relaxalg} with incumbent $\mathbf{S}_{M_o}^{\circ}=\mathbf{S}_{M_o}^{(i)}$.}
    \State \multiline{Fix $\mathbf{S}_{M_o}^{(i+1)}$ and update $\boldsymbol{\Phi}^{(i+1)}$ via Algorithm~\ref{adp} with incumbent $\boldsymbol{\Phi}^{\circ}=\boldsymbol{\Phi}^{(i)}$.}
    \State Compute $\mathcal{J}^{(i+1)} = \mathcal{J}(\mathbf{S}_{M_o}^{(i+1)}, \boldsymbol{\Phi}^{(i+1)})$.
    \State $i \leftarrow i+1$
\EndWhile
\State \Return $(\boldsymbol{\Phi}^{(i)}, \mathbf{S}_{M_o}^{(i)})$.
\end{algorithmic}
\end{algorithm}
\subsection{Computational Complexity}\label{compcomp}
We summarize the complexity of the proposed frameworks by their dominant terms. In Algorithm~\ref{CEM}, each of the $T_{\mathrm{cem}}$ iterations evaluates $\mathcal{J}$ for $N_{\mathrm{s}}$ samples via matrix-vector products in $\mathcal{O}(N_{\mathrm{s}}M^2)$, while sampling, budget correction, and elite selection require $\mathcal{O}(N_{\mathrm{s}}M\log M+N_{\mathrm{s}}\log N_{\mathrm{s}})$; hence, its complexity is $\mathcal{O}\big(T_{\mathrm{cem}}(N_{\mathrm{s}}M^2+N_{\mathrm{s}}\log N_{\mathrm{s}})\big)$. Algorithm~\ref{relaxalg} is dominated by solving the SDP~\eqref{eq:CC_SDR_link} via an interior-point method, whose iterations each cost $\mathcal{O}(M^6)$ owing to the $\mathcal{O}(M^2)$ element-wise constraints~\cite{sdr, boyd}, yielding $\mathcal{O}(T_{\mathrm{ip}}M^6)$ for $T_{\mathrm{ip}}$ iterations; the subsequent randomization, rounding, and evaluation of $N_{\mathrm{r}}$ candidates require only $\mathcal{O}(M^3+N_{\mathrm{r}}M^2)$. In Algorithm~\ref{adp}, each Riemannian step with $N_{\mathrm{ls}}$ Armijo backtracking evaluations costs $\mathcal{O}(N_{\mathrm{ls}}M_o^2)$, leading to $\mathcal{O}(LT_{\mathrm{riem}}N_{\mathrm{ls}}M_o^2)$ for $T_{\mathrm{riem}}$ inner and $L$ outer iterations. Consequently, over $I_1$ and $I_2$ AO iterations, the total complexities of AO-CEM and AO-SDR are given by
\begin{equation}
\label{comao1}
\begin{aligned}
&\mathcal{O}\Big(I_1\big(T_{\mathrm{cem}}(N_{\mathrm{s}}M^2+N_{\mathrm{s}}\log N_{\mathrm{s}})+LT_{\mathrm{riem}}N_{\mathrm{ls}}M_o^2\big)\Big),\\
&\mathcal{O}\Big(I_2\big(T_{\mathrm{ip}}M^6+LT_{\mathrm{riem}}N_{\mathrm{ls}}M_o^2\big)\Big),
\end{aligned}
\end{equation}
respectively, where the $\mathcal{O}(M^2)$ cost of evaluating $\mathcal{J}^{(i+1)}$ is absorbed. Since $M^6\gg M^2$, AO-CEM is substantially lighter than AO-SDR, whereas AO-SDR achieves slightly better performance, as will be demonstrated in Section~\ref{roro}.

%
%
%
%

{\section{Ergodic Capacity Enhancement by the Proposed AO Framework}\label{erca}
The ergodic capacity of the considered FRIS-assisted communication system is defined as follows:
\begin{equation}
\label{ergcap}
C_{\mathrm{erg}} \triangleq \mathbb{E}[\log_2 (1+\gamma L_f L_u |\mathbf{H}_{\text{eq}}|^2 )],
\end{equation}
\begin{table}[t]
\centering
\caption{Simulation Parameters}
\label{tabsim}
\begin{tabular}{|>{\centering}m {5.8cm}|>{\centering}m {2cm}|}
\hline
\textbf{Parameter} & \textbf{Value} 
\tabularnewline
\hline
Target rate $R$ & $1$ [bps/Hz] 
\tabularnewline
\hline
Number of FRIS elements $M$ (unless referred) & 100~($10\times 10$)
\tabularnewline
\hline
Transmit SNR $\bar{\gamma}$ (unless referred) & 15 [dB]
\tabularnewline
\hline
Normalized FRIS aperture $W_x$ (unless referred) & 2
\tabularnewline
\hline
PLE $\alpha$, Rician factor $K$, and Carrier frequency $f_c$ & $(2.2,~1,~5~[\text{GHz}])$ 
\tabularnewline
\hline
\end{tabular}
\end{table}
Consider any Pareto improvement~\cite{multio}:
\begin{equation}
\label{paretoimpro}
\left(\lambda',\sigma^{2\prime}\right) \succ \left(\lambda,\sigma^2\right)(\lambda'\ge\lambda,\sigma^{2\prime}\ge\sigma^2~\text{and at least one strict}),
\end{equation}
which reduces $P_{\text{out}}=1-Q_1\left(\sqrt{2\lambda}, \sqrt{\frac{c}{\sigma^2}}\right)$ with $c$ in~(\ref{lamsig}). For exact ordering of $C_{\rm erg}$ under~\eqref{paretoimpro}, consider two fixed feasible FRIS configurations with equivalent $(X(=|\mathbf{H}_{\mathrm{eq}}|^2), X')$, $(P_{\rm out}, P_{\rm out}')$ and $(C_{\rm erg}, C_{\rm erg}')$, where primed quantities refer to the second configuration. Since $\mathbb P(X>t)=Q_1\big(\sqrt{2\lambda},\sqrt{2t/\sigma^2}\big)$ by~(\ref{meanva}) and~(\ref{outr}), and $Q_1(a,b)$ is increasing in $a$ and decreasing in $b$~\cite{marQ},~\eqref{paretoimpro} yields the ordering:
\begin{equation}
\label{sfo}
\mathbb P(X'>t)\ge\mathbb P(X>t)~(\forall t\ge0),
\end{equation}
with strict inequality for every $t>0$. Thus, $X'$ first-order stochastically dominates $X$; in particular,~\eqref{sfo} reduces the outage probability at every target rate simultaneously, not only at the design rate $R$. Thereafter, the exact $C_{\mathrm{erg}}$ admits:
\begin{equation}
\label{eq:capacity_tail}
C_{\mathrm{erg}}
=
\frac{1}{\ln 2}
\int_0^\infty
\frac{\eta \Pr(X>t)}{1+\eta t} dt~(\eta\triangleq\gamma L_fL_u).
\end{equation}
The ordering~\eqref{sfo} therefore implies $C'_{\mathrm{erg}}>C_{\mathrm{erg}}$ under~\eqref{paretoimpro}.\footnote{In essence, increasing both $\lambda$ and $\sigma^2$ pushes the effective channel distribution ($|\mathbf H_{\mathrm{eq}}|^2$) further away from the outage region. Consequently, fewer fading realizations fall below the target threshold, improving reliability, and more realizations contribute constructively to $C_{\rm erg}$. This indicates that reliability and throughput can be simultaneously boosted by steering ($\lambda,\sigma^2$) in a favorable direction.} Hence, reliability and capacity are improved simultaneously by any Pareto improvement of $(\lambda,\sigma^2)$, which motivates the $\epsilon$-constraint design of the subproblems in Sections~\ref{sec:gradient_relaxation} and~\ref{phi}, where $M_1$ and $M_2$ are set to the power of the incumbent so that an exact solution of each subproblem would not decrease either parameter.
\begin{remark}
\label{remacc}
An increase of $\mathcal{J}$ at $R$ alone, i.e., a decrease of the CDF of $X$ at a single threshold, does not imply~\eqref{sfo}, and $C_{\rm erg}$ may then decrease. Hence, the simultaneous improvement of $P_{\rm out}$ and $C_{\rm erg}$ is guaranteed for Pareto improvements of $(\lambda,\sigma^2)$, rather than for any reduction of $P_{\rm out}$ at the design rate. 
\end{remark}
The simultaneous improvements of both metrics by the proposed AO frameworks are numerically demonstrated in Section~\ref{numer}.
}

\section{Numerical Results}\label{numer}
In our simulations, we evaluate the validity of the proposed AO frameworks for $P_{\text{out}}$ and ${C}_{\mathrm{erg}}$ by means of Monte-Carlo simulations. We adopt an FRIS system with the three-dimensional (3D) coordinate of BS, FRIS, and MU with $l_f=4~\mathrm{m}$ and $l_u=16~\mathrm{m}$. 
Herein, $l_f =4~\mathrm{m}$ and $l_u=16~\mathrm{m}$ exceed $D^{\mathrm{R}}=4W_x^2\lambda_c\approx0.96~\mathrm{m}$ for the default $W_x=2$ at $f_c=5~\mathrm{GHz}$, so that the planar-wavefront LoS model is accurate~\cite{neardai, nearris}. For the larger apertures in Fig.~\ref{fig_osize}, $D^{\mathrm{R}}$ exceeds $l_f$ when $W_x\gtrapprox4.1$; we retain the planar-wavefront model there for comparability, and the near-field extension is left for future work. 
The angles of the LoS components follow $\mathcal{U}[0, \pi]$, and 
the detailed simulation parameters are summarized in Table~\ref{tabsim}. 

We benchmark the optimality against the BFS that exhaustively enumerates all possible port-phase configurations of FRIS with assuming $b=2$-bits discrete phase shift~\cite{FRISonoff}. Note that the BFS yields the optimum of~(\ref{objoriginal}) over the $b$-bit discrete phase set, which serves as a practical near-optimal benchmark rather than the global optimum of~(\ref{objoriginal}) with continuous phase shifts. For both $P_{\text{out}}$ and $C_{\mathrm{erg}}$ evaluations, we consider a conventional RIS system of the same $M_o$ elements, and additionally adopt the E-PSO~\cite{VegaSanchez2026} and the CEM-based design~\cite{FRISonoff} for $P_{\text{out}}$- and $C_{\mathrm{erg}}$-comparison, denoted by \textbf{E-PSO} and \textbf{CEM}, respectively.


\subsection{Reliability and Optimality of Proposed AO Frameworks}\label{roro}
\begin{figure}[t]
	\begin{center}
		\includegraphics[width=0.95\columnwidth,keepaspectratio]%
		{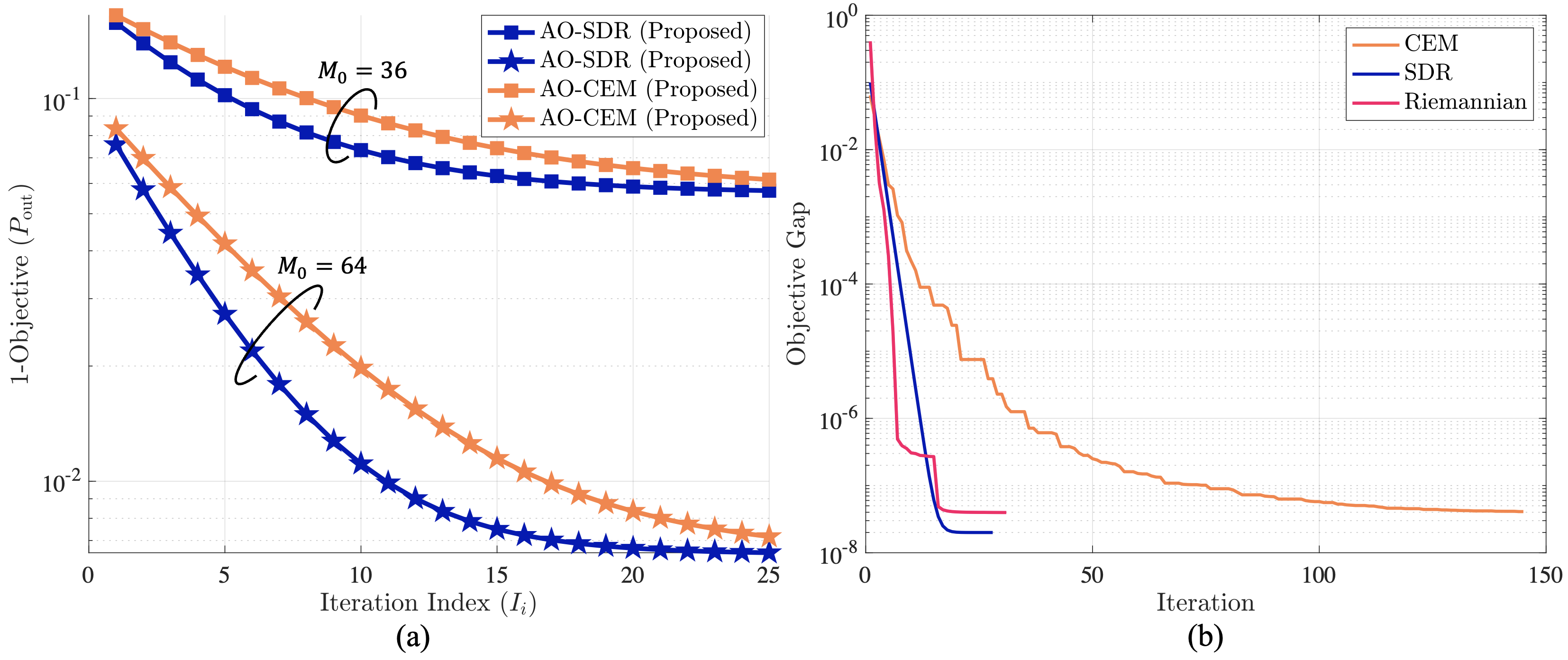}
		\caption{Convergence performance of (a) Algorithm~\ref{aocem} and~\ref{aorelax} for different $M_o$ and (b) Algorithm~\ref{CEM}$\sim$~\ref{adp} with $M_o=36$.}
		\label{fig_conv}
	\end{center}
\end{figure}

Fig.~\ref{fig_conv} presents the convergence characteristics of (a) the proposed AO-SDR and AO-CEM schemes under different values of $M_o$ and (b) the inner frameworks (SDR, CEM and Riemannian), assessed in terms of the objective value and gap across iterations, respectively. In both, the algorithms exhibit improvement and typically reach convergence within about 20 iterations, highlighting their practicality for implementation. In particular, in (b), the SDR- and Riemannian-based methods require fewer iterations, since the SDR is solved by an interior-point method whose duality gap decreases rapidly and the Riemannian ascent exploits gradient information on the unit-modulus manifold, whereas CEM-based method performs lightweight stochastic updates based on sampling and elite selection, which appears considerably slower in convergence plots. Nevertheless, this limitation of CEM is compensated by its superior computational complexity compared to SDR-based measured by realistic computation time, as analytically shown in Section~\ref{compcomp} and will be numerically confirmed later. Moreover, as illustrated in (a), increasing $M_o$ slightly slows down convergence since the expanded search space, coupled with stronger spatial correlation, reduces the marginal improvement gained per $\{I_m\}$~\cite{curse}.
\begin{figure}[t]
	\begin{center}
		\includegraphics[width=0.98\columnwidth,keepaspectratio]%
		{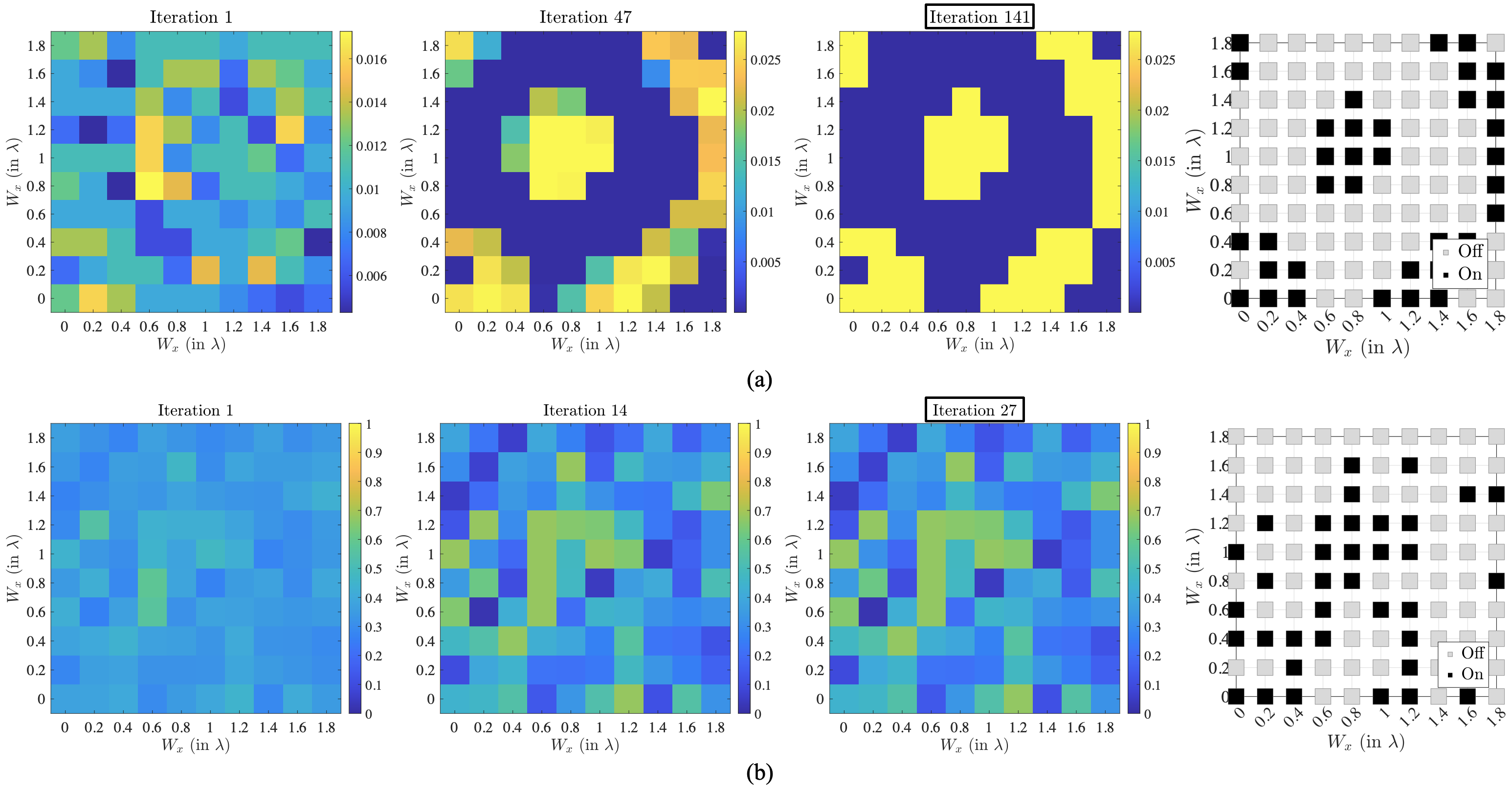}
		\caption{Evolution of port selection probabilities over iterations and final (optimal) FRIS activation patterns with (a) CEM and (b) SDR-based framework, where $M_o=36$.}
		\label{fig_ex}
	\end{center}
\end{figure}

Fig.~\ref{fig_ex}(a) and (b) illustrate the port selection processes of the proposed CEM and SDR-based framework in Algorithm~\ref{CEM} and~\ref{relaxalg}, respectively, for $M_o=36$. In both cases, the procedure begins with an almost uniform distribution, which gradually concentrates probability mass onto specific ports. This convergence behavior highlights how the probability updates (in CEM) and the relaxation-and-refinement mechanism (in SDR) progressively enforce sparsity in the port selection process, where CEM starts from an almost uniform distribution and gradually sharpens the probabilities, whereas SDR directly derives the decision from a continuous-valued $\mathbf w$. Furthermore, consistent with Fig.~\ref{fig_conv}(b), the required iteration number for SDR is observed to be smaller than that of CEM.

Table~\ref{matime} presents the MATLAB execution times of the proposed AO-SDR and AO-CEM frameworks, along with those of the BFS for the case of $M=16$ and $M_o \in \{4,6\}$. This evaluation complements the theoretical complexity discussion in Section~\ref{compcomp}, while also providing a practical benchmark against BFS. As observed, AO-SDR and AO-CEM require about $15$ and less than $2.5\%$ of the BFS computation time for $M_o=4$ and under $6$ and $1\%$ for $M_o=6$. This dramatic complexity reduction occurs since BFS exhaustively enumerates all possible port-phase configurations of FRIS, whose search space grows combinatorially with $M_o$ and $b$. In contrast, the targeted search of the proposed AO frameworks avoids exhaustive enumeration of non-optimal configurations, thereby requiring far fewer evaluations and yielding significantly lower runtimes than BFS. These findings demonstrate that the proposed AO frameworks offer significant computational savings not only from an analytical (big-$\mathcal{O}$) perspective but also in terms of actual execution time. Furthermore, AO-CEM achieves approximately $13\sim16\%$ of the runtime of AO-SDR, in line with the analysis in Section~\ref{compcomp} and Fig.~\ref{fig_conv}(a) that CEM is superior in terms of computational complexity compared to SDR with similar scale of $\{I_m\}$ with SDR, thereby validating the practical complexity advantage. It is worth noting, however, that Table~\ref{matime} should be regarded as a rough indicator of relative computational complexity.
\begin{table}[t]
\centering
\caption{Average MATLAB Computation Time $(M=16)$}
\begin{tabular}{|c|c|c|c|}
\hline
$\mathbf{M_o}$ & \textbf{Method} & \textbf{Time (sec)} & \textbf{Relative to BFS} \\
\hline
\multirow{3}{*}{4} 
 & BFS  & 1.475 & 1 \\
 & \textbf{AO-SDR (Proposed)}  & 0.226 & $15.32\%$ \\
 & \textbf{AO-CEM (Proposed)}  & 0.036 & $2.44\%$ \\
\hline
\multirow{3}{*}{6} 
 & BFS     & 5.484 & 1.00 \\
 & \textbf{AO-SDR (Proposed)}  & 0.307 & $5.60\%$ \\
 & \textbf{AO-CEM (Proposed)} & 0.041 & $ 0.75\%$ \\
\hline
\end{tabular}
\label{matime}
\end{table}
\begin{figure}[t]
	\begin{center}
		\includegraphics[width=0.95\columnwidth,keepaspectratio]%
		{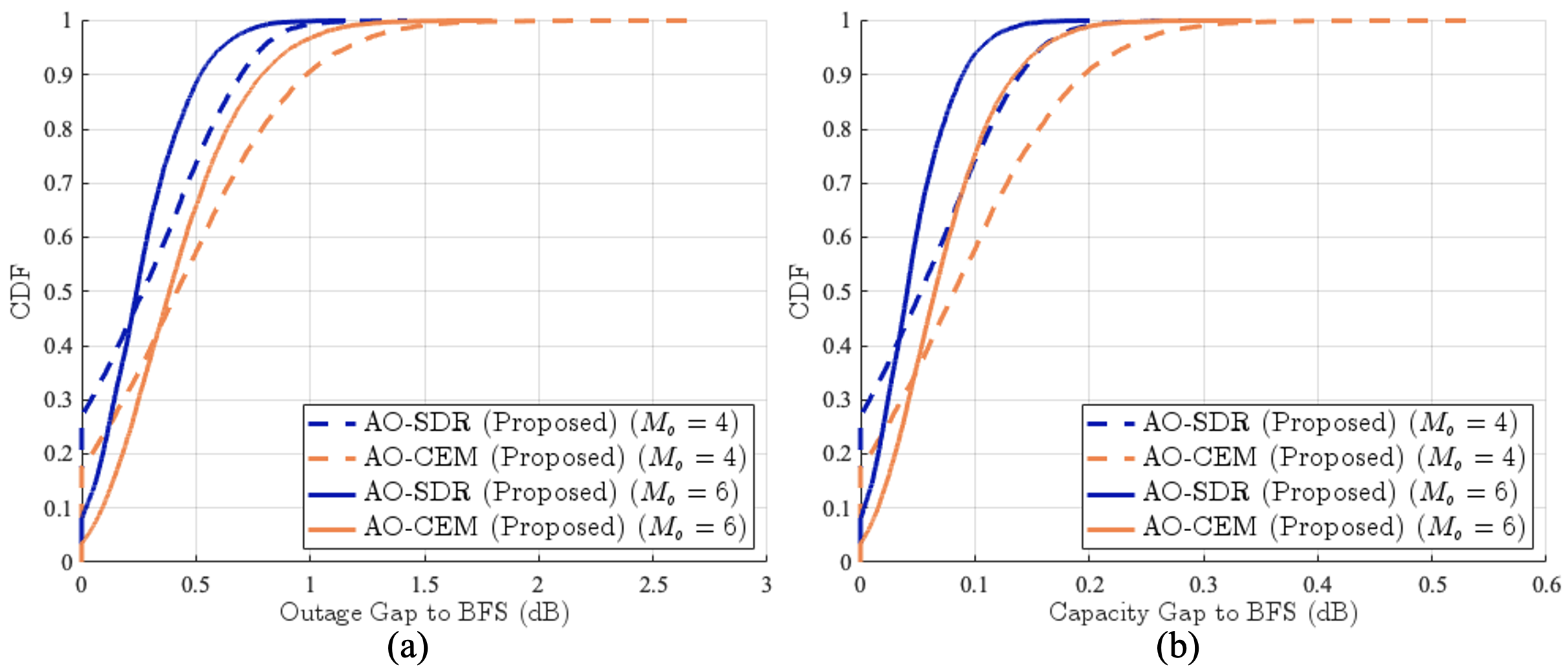}
		\caption{CDF of (a) outage probability and (b) ergodic capacity gap between BFS and proposed AO frameworks for $M=16$ and different $M_o$.}
		\label{fig_cdfout}
	\end{center}
\end{figure}
\begin{table}[t]
\centering
\caption{Mean gap to BFS for outage probability and ergodic capacity in Fig.~\ref{fig_cdfout} $(M=16)$}
\label{table_gap}
\begin{tabular}{|c|cc|cc|}
\hline
\multirow{2}{*}{$\mathbf{M_o}$} & \multicolumn{2}{c|}{\textbf{Gap of} $\mathbf{P_{\text{out}}}$ \textbf{(dB)}} & \multicolumn{2}{c|}{\textbf{Gap of} $\mathbf{C_{\mathrm{erg}}}$ \textbf{(dB)}} \\
\cline{2-5}
 & \textbf{AO-SDR} & \textbf{AO-CEM} & \textbf{AO-SDR} & \textbf{AO-CEM} \\
\hline
$4$ & $0.5406$ & $0.6976$ & $0.0602$ & $0.0910$ \\
$6$ & $0.5816$ & $0.7374$ & $0.0448$ & $0.0713$ \\
\hline
\end{tabular}
\end{table}

To quantitatively assess the insignificance of the optimality gap, we evaluated the CDFs of $P_{\text{out}}$ and $C_{\mathrm{erg}}$ gaps between the BFS and the proposed AO framework through Monte Carlo simulations. Fig.~\ref{fig_cdfout}(a) and the left panel of Table~\ref{table_gap} show the CDF and the mean outage probability gap between the BFS benchmark and the proposed AO frameworks for $M=16$ with $M_o \in \{4,6\}$. The findings confirm that the AO-based solutions operate close to optimal, with the largest mean gap under $0.74~\mathrm{dB}$, which is negligible. In addition, AO-SDR consistently achieves a smaller gap than AO-CEM, underscoring its advantage in outage performance.

Similarly, Fig.~\ref{fig_cdfout}(b) and the right panel of Table~\ref{table_gap} depict the CDF and the mean ergodic capacity gap under the same conditions. The proposed AO frameworks again demonstrate near-optimality, with the maximum mean gap approximately below $0.09~\mathrm{dB}$, which is practically insignificant. Therein, AO-SDR again exhibits a smaller performance gap than AO-CEM, underscoring its advantage in terms of capacity. From Fig.~\ref{fig_cdfout}(b), the proposed AO frameworks are shown to also enhance ergodic capacity while maintaining a small optimality gap.
\subsection{Outage Probability and Ergodic Capacity Comparisons}
\begin{figure}[t]
	\begin{center}
		\includegraphics[width=0.95\columnwidth,keepaspectratio]%
		{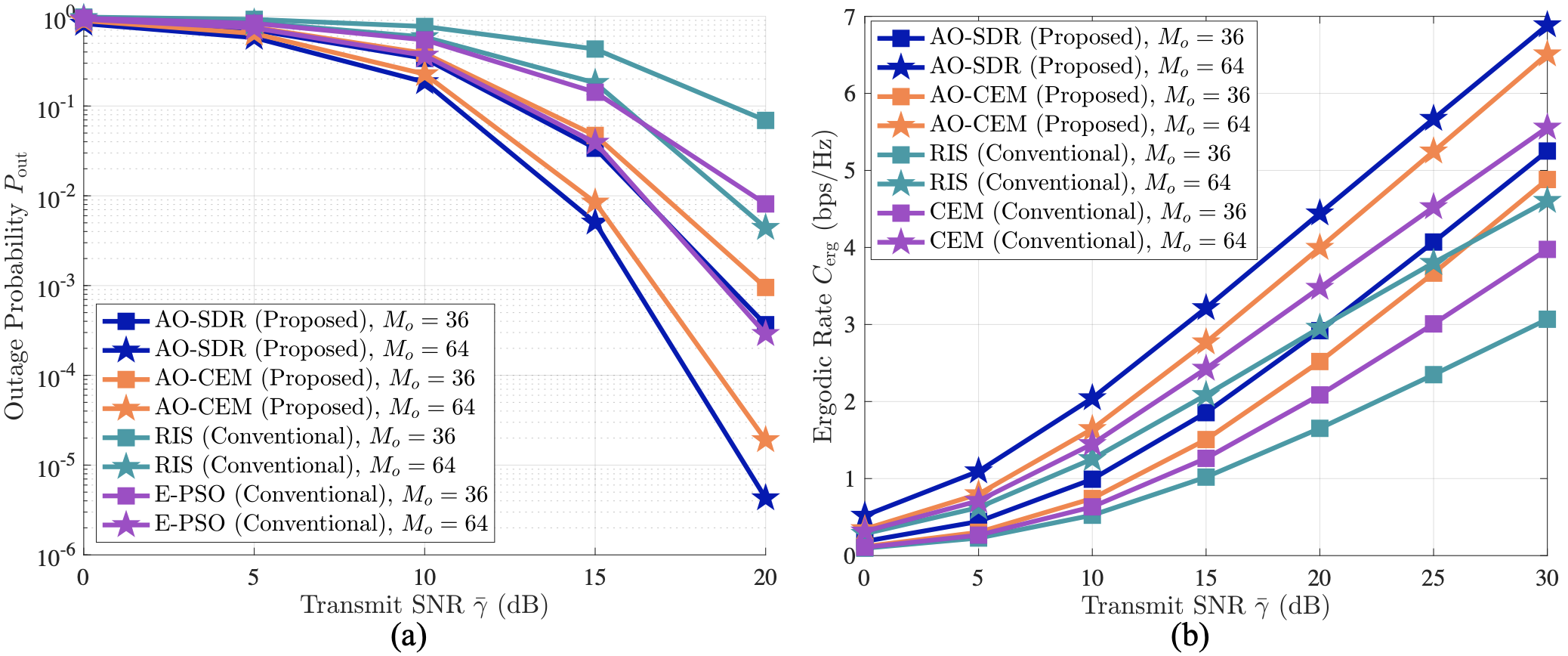}
		\caption{(a) Outage probability $P_{\text{out}}$ and (b) Ergodic Capacity $C_{\mathrm{erg}}$ versus the transmit SNR $\bar{\gamma}$ for different $M_o$.}
		\label{fig_outage}
	\end{center}
\end{figure}

Fig.~\ref{fig_outage}(a) illustrates $P_{\mathrm{out}}$ versus $\bar{\gamma}$ for different $M_o$. The proposed AO-SDR and AO-CEM frameworks achieve lower $P_{\mathrm{out}}$ than both the conventional RIS and E-PSO benchmarks under the same active-element budget. The performance separation becomes particularly pronounced at high SNR, indicating that the proposed designs more effectively reduce the probability of weak equivalent-channel realizations. For example, at $\bar{\gamma}=20~\mathrm{dB}$, AO-SDR achieves $P_{\mathrm{out}}$ of approximately $4\times10^{-4}$ and $4\times10^{-6}$ for $M_o=36$ and $M_o=64$, respectively. These values represent reductions of roughly two and three orders of magnitude relative to conventional RIS. Although E-PSO also improves upon conventional RIS, both proposed frameworks provide further reductions on $P_{\mathrm{out}}$, with a larger separation observed for $M_o=64$ at high SNR. Increasing $M_o$ from 36 to 64 further improves reliability, highlighting the benefit of configuring a larger number of $M_o$. Moreover, AO-SDR consistently outperforms AO-CEM across the entire SNR range, owing to the deterministic convex-relaxation framework of SDR, which efficiently drives rapid gap reduction and reliably identifies high-quality FRIS port-phase configurations, thereby providing a small but consistent performance advantage.

Fig.~\ref{fig_outage}(b) shows $C_{\mathrm{erg}}$ versus $\bar{\gamma}$ for the evaluated schemes with different $M_o$. Both proposed AO-SDR and AO-CEM frameworks achieve higher $C_{\mathrm{erg}}$ than the conventional RIS and CEM benchmarks, particularly in the moderate-to-high SNR regime. These improvements complement the reductions on $P_{\text{out}}$ observed in Fig.~\ref{fig_outage}(a), demonstrating gains in both reliability and spectral efficiency. Although the CEM benchmark outperforms conventional RIS, the proposed frameworks provide further improvements on $C_{\mathrm{erg}}$, with the absolute performance gaps increasing over the plotted SNR range. For $M_o=36$, AO-SDR achieves gains on $C_{\mathrm{erg}}$ of approximately $0.8$ and $1.3~\mathrm{bps/Hz}$ over conventional RIS at $\bar{\gamma}=15$ and $20~\mathrm{dB}$, respectively. For $M_o=64$, the corresponding gain reaches approximately $2.3~\mathrm{bps/Hz}$ at $\bar{\gamma}=20~\mathrm{dB}$. At $\bar{\gamma}=30~\mathrm{dB}$, AO-SDR with $M_o=64$ achieves approximately $6.8~\mathrm{bps/Hz}$, compared with $5.5~\mathrm{bps/Hz}$ for CEM and $4.6~\mathrm{bps/Hz}$ for conventional RIS. Notably, AO-SDR with $M_o=36$ approximately matches the performance of conventional RIS with $M_o=64$, indicating more efficient use of selected FRIS elements. Moreover, AO-SDR maintains an advantage on $C_{\mathrm{erg}}$ over AO-CEM in the considered scenarios, consistent with the outage comparison in Fig.~\ref{fig_outage}(a).
\begin{figure}[t]
	\begin{center}
		\includegraphics[width=0.95\columnwidth,keepaspectratio]%
		{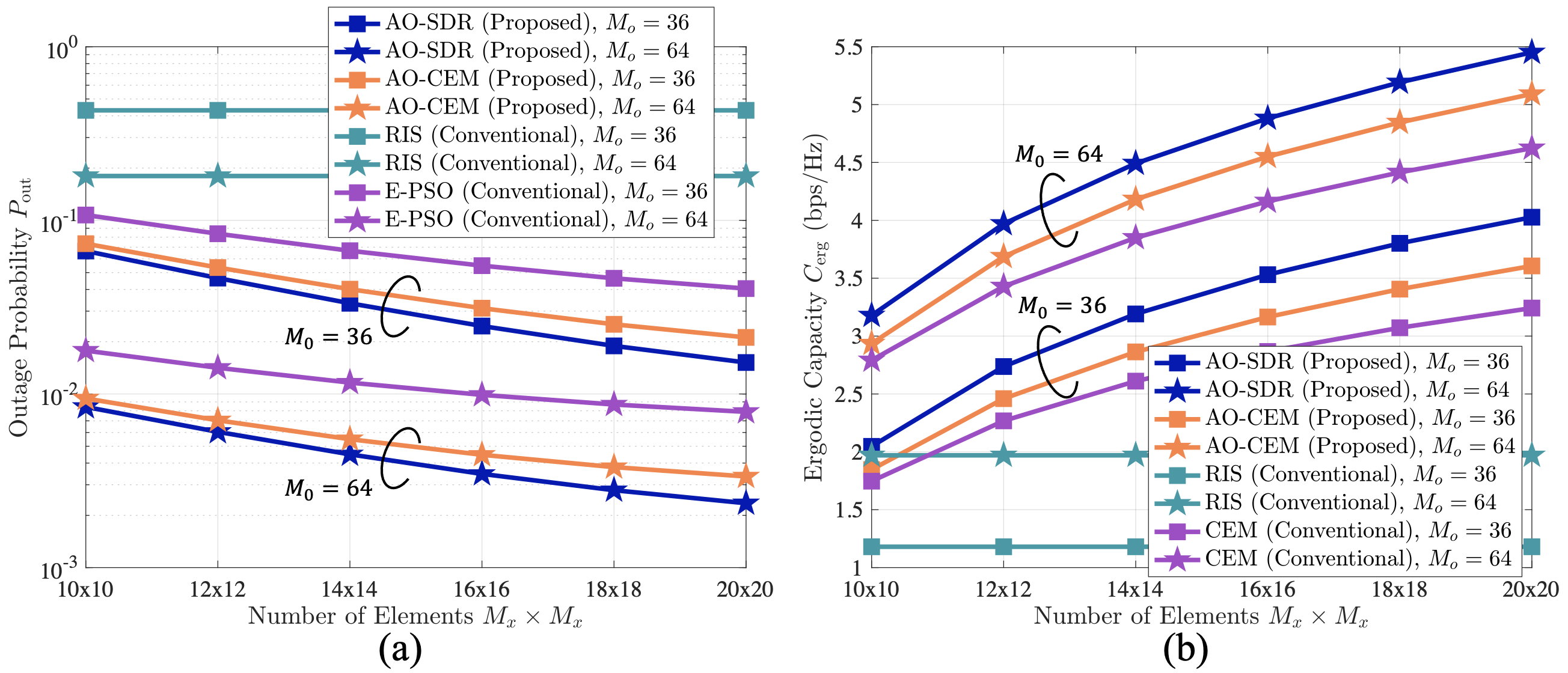}
		\caption{(a) Outage probability $P_{\text{out}}$ and (b) Ergodic Capacity $C_{\mathrm{erg}}$ versus the number of FRIS elements $M$ for different $M_o$.}
		\label{fig_onum}
	\end{center}
\end{figure}

Fig.~\ref{fig_onum}(a) presents $P_{\mathrm{out}}$ as a function of $M=M_x\times M_x$ for different numbers of $M_o$. Increasing $M$ provides more candidate port configurations while keeping $M_o$ fixed, resulting in a consistent reduction on $P_{\mathrm{out}}$ for both proposed AO-SDR and AO-CEM frameworks. Both methods outperform the conventional RIS and E-PSO benchmarks throughout the considered range. Although E-PSO also benefits from the larger candidate set, the proposed frameworks achieve further reliability improvements through more effective port-phase configurations. In contrast, the conventional RIS curves remain unchanged since its fixed element configuration does not exploit the additional candidate ports. At $M=20\times20$, AO-SDR achieves $P_{\mathrm{out}}$ of approximately $1.5\times10^{-2}$ and $2.3\times10^{-3}$ for $M_o=36$ and $M_o=64$, respectively, corresponding to approximately 30- and 80-fold reductions relative to conventional RIS. AO-SDR also reduces $P_{\mathrm{out}}$ by roughly a factor of three relative to E-PSO for both values of $M_o$. Moreover, increasing $M_o$ from 36 to 64 substantially lowers the $P_{\mathrm{out}}$, while AO-SDR maintains a performance advantage over AO-CEM. These results highlight the benefits of optimizing port activation and reflection phases over an expanded set of candidate configurations.
\begin{figure}[t]
	\begin{center}
		\includegraphics[width=0.95\columnwidth,keepaspectratio]%
		{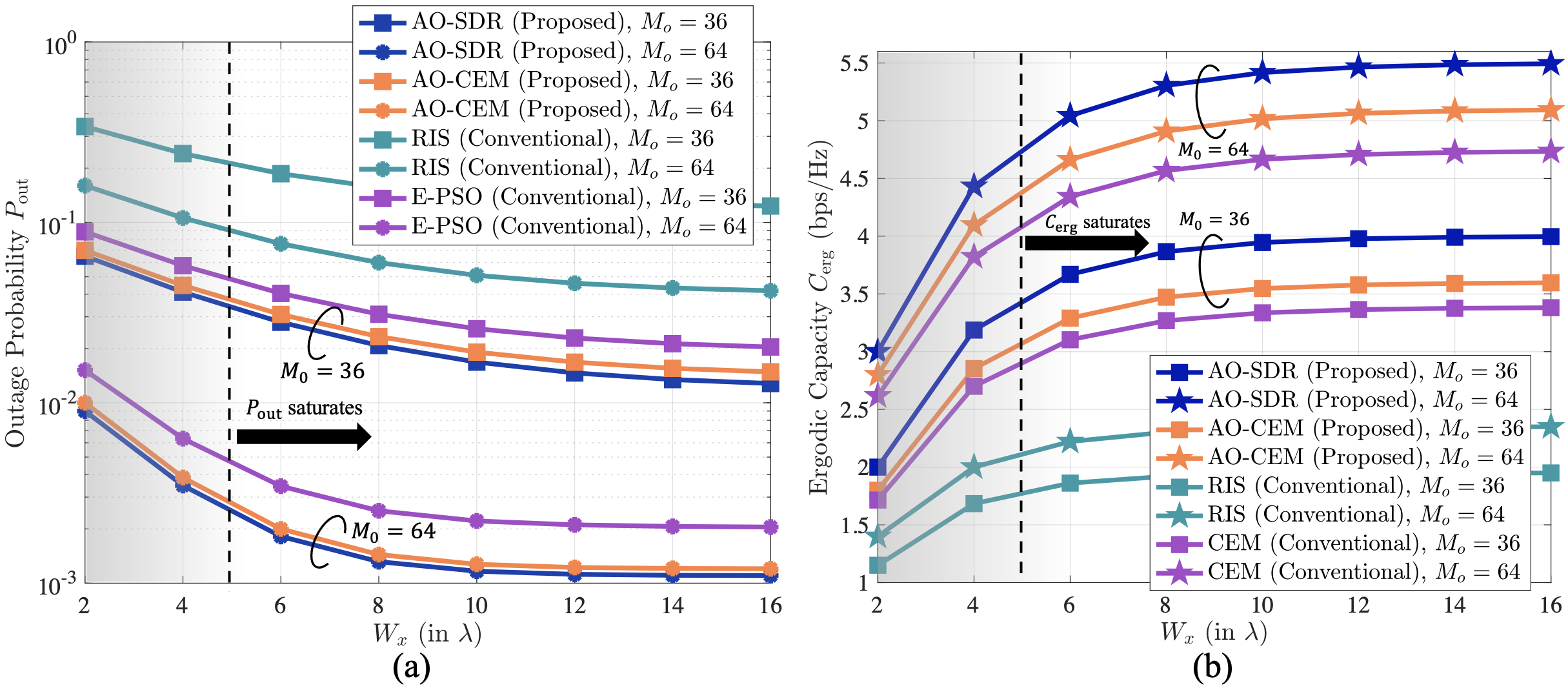}
		\caption{(a) Outage probability $P_{\text{out}}$ and (b) Ergodic Capacity $C_{\mathrm{erg}}$ versus the aperture $W_x$ for different $M_o$.}
		\label{fig_osize}
	\end{center}
\end{figure}

Fig.~\ref{fig_onum}(b) presents $C_{\mathrm{erg}}$ as a function of $M=M_x\times M_x$ for different $M_o$. Both proposed AO-SDR and AO-CEM frameworks consistently outperform the conventional RIS and CEM benchmarks. Increasing $M$ expands the set of available port configurations while keeping the number of active reflecting elements fixed, enabling further improvements on $C_{\mathrm{erg}}$ through port and phase optimization. The CEM benchmark also benefits from this increased selection flexibility, whereas conventional RIS maintains a constant $C_{\mathrm{erg}}$ since its fixed configuration does not exploit the additional candidate ports. At $M=20\times20$, AO-SDR achieves approximately $4.0$ and $5.5~\mathrm{bps/Hz}$ for $M_o=36$ and $M_o=64$, respectively, compared with $3.2$ and $4.6~\mathrm{bps/Hz}$ for CEM and $1.2$ and $2.0~\mathrm{bps/Hz}$ for conventional RIS. The corresponding gains over conventional RIS are approximately $2.8$ and $3.5~\mathrm{bps/Hz}$, highlighting the increasing advantage of the proposed design as the candidate set grows. Notably, even at $M=10\times10$, AO-SDR with $M_o=36$ achieves $C_{\mathrm{erg}}$ comparable to conventional RIS with $M_o=64$, demonstrating more efficient use of active reflecting elements. Moreover, AO-SDR maintains an advantage on $C_{\mathrm{erg}}$ over AO-CEM throughout the considered range, consistent with the outage results in Fig.~\ref{fig_onum}(a).

Fig.~\ref{fig_osize}(a) presents $P_{\mathrm{out}}$ as a function of $W_x$ for different $M_o$. Increasing $W_x$ expands the spatial aperture, enabling more effective port selection and thereby reducing channel correlation, which in turn enhances performance on $P_{\text{out}}$. Both proposed AO-SDR and AO-CEM frameworks achieve lower $P_{\text{out}}$ than the conventional RIS and E-PSO benchmarks throughout the considered range. Although E-PSO also benefits from aperture expansion, the proposed frameworks provide further reliability improvements. At $W_x=16$, AO-SDR achieves $P_{\text{out}}$ of approximately $1.3\times10^{-2}$ and $1.1\times10^{-3}$ for $M_o=36$ and $M_o=64$, respectively, compared with approximately $1.2\times10^{-1}$ and $4.2\times10^{-2}$ for conventional RIS. These values correspond to approximately $10$-fold and $40$-fold reductions on $P_{\text{out}}$. Beyond the range of $W_x\gtrapprox 5~(\because M_x=10)$~\cite{corrhalf,newcor, perlimfas}, however, all curves exhibit saturation; further increases in aperture produce only minor improvements. This saturation arises since the spatial correlation between ports, characterized by the spherical Bessel function, which experiences its steepest decline around $\frac{\lambda_c}{2}$~\cite{perlimfas, bessbook}, is already sufficiently mitigated when $W_x > 5$ (i.e., $d > \frac{\lambda_c}{2}$). Beyond this point, further increasing $W_x$ yields only marginal diversity gains. Notably, the proposed AO frameworks leverage this transitional regime more effectively than benchmarks.

Fig.~\ref{fig_osize}(b) depicts $C_{\mathrm{erg}}$ as a function of $W_x$ for different $M_o$. Both proposed AO-SDR and AO-CEM frameworks achieve higher $C_{\mathrm{erg}}$ than the conventional RIS and CEM benchmarks. The improvements on $C_{\mathrm{erg}}$ are particularly pronounced over the smaller-aperture range, consistent with reduced spatial correlation and improved opportunities for selecting favorable port configurations. For example, increasing $W_x$ from $2$ to $4$ raises the AO-SDR rate by approximately $1.2$ and $1.4~\mathrm{bps/Hz}$ for $M_o=36$ and $M_o=64$, respectively. The corresponding improvements for AO-CEM are approximately $1.1$ and $1.3~\mathrm{bps/Hz}$. Consistent with Fig.~\ref{fig_osize}(a), as $W_x\gtrapprox5$, the capacity gain begins to saturate since additional aperture offers only marginal decorrelation benefits~\cite{perlimfas}. Notably, the proposed frameworks exploit this transitional regime more efficiently than benchmarks.

From Figs.~\ref{fig_outage}$\sim$\ref{fig_osize}, we can see that AO-SDR outperforms AO-CEM, and combining this with AO-CEM's lower complexity derived both analytically in Section~\ref{compcomp} and numerically in Table~\ref{matime}, we can determine the trade-off between the complexity and the performance of the proposed AO-SDR and AO-CEM frameworks. Furthermore, Figs.~\ref{fig_onum}$\sim$\ref{fig_osize} indicate that $M$ has a more pronounced influence than $W_x$ on FRIS performance with saturation, consistent with the trends previously reported for FAS~\cite{perlimfas, FAS}.
\section{Conclusion}
In this paper, we proposed an FRIS framework that jointly optimizes port selection and phase shifts to improve link reliability and spectral efficiency. Two AO frameworks were developed: AO-CEM, which explores the combinatorial port space via probabilistic elite-sample updates, and AO-SDR, which identifies high-quality port configurations via convex relaxation, both combined with a Dinkelbach-based Riemannian method with penalty for phase optimization on the unit-modulus manifold. We further showed that Pareto improvements in the equivalent channel parameters simultaneously reduce the outage probability and increase the ergodic capacity, which motivates the $\epsilon$-constraint design of the proposed subproblems. Simulations demonstrated significant gains over the baselines, with AO-SDR achieving the best performance and AO-CEM offering comparable performance at lower complexity, both attaining near-optimal performance relative to BFS at a fraction of its runtime.

		\bibliographystyle{IEEEtran}
		\bibliography{IEEEexample}	

\begin{thebibliography}{10}
\providecommand{\url}[1]{#1}
\csname url@samestyle\endcsname
\providecommand{\newblock}{\relax}
\providecommand{\bibinfo}[2]{#2}
\providecommand{\BIBentrySTDinterwordspacing}{\spaceskip=0pt\relax}
\providecommand{\BIBentryALTinterwordstretchfactor}{4}
\providecommand{\BIBentryALTinterwordspacing}{\spaceskip=\fontdimen2\font plus
\BIBentryALTinterwordstretchfactor\fontdimen3\font minus
  \fontdimen4\font\relax}
\providecommand{\BIBforeignlanguage}[2]{{%
\expandafter\ifx\csname l@#1\endcsname\relax
\typeout{** WARNING: IEEEtran.bst: No hyphenation pattern has been}%
\typeout{** loaded for the language `#1'. Using the pattern for}%
\typeout{** the default language instead.}%
\else
\language=\csname l@#1\endcsname
\fi
#2}}
\providecommand{\BIBdecl}{\relax}
\BIBdecl

\bibitem{RISvtm}
E.~Basar \emph{et~al.}, ``Reconfigurable intelligent surfaces for {6G}:
  Emerging hardware architectures, applications, and open challenges,''
  \emph{IEEE Veh. Technol. Mag.}, vol.~19, no.~3, pp. 27--47, Sep. 2024.

\bibitem{RIST}
Q.~Wu \emph{et~al.}, ``Intelligent reflecting surface-aided wireless
  communications: A tutorial,'' \emph{IEEE Trans. Commun.}, vol.~69, no.~5, pp.
  3313--3351, May 2021.

\bibitem{holotut}
T.~Gong \emph{et~al.}, ``Holographic {MIMO} communications: Theoretical
  foundations, enabling technologies, and future directions,'' \emph{{IEEE}
  Commun. Surveys Tuts.}, vol.~26, no.~1, pp. 196--257, First quarter 2024.

\bibitem{dsRIS}
D.~Jun \emph{et~al.}, ``Reconfigurable intelligence surface with potential
  tunable meta-devices for {6G}: Design and system-level evaluation,''
  \emph{IEEE Commun. Stand. Mag.}, vol.~8, no.~4, pp. 32--39, Apr. 2024.

\bibitem{HBRIS}
H.-B. Jeon \emph{et~al.}, ``An energy-efficient aerial backhaul system with
  reconfigurable intelligent surface,'' \emph{IEEE Trans. Wireless Commun.},
  vol.~21, no.~8, pp. 6478--6494, Aug. 2022.

\bibitem{tkz}
T.~Wu \emph{et~al.}, ``{XL-RIS}-aided near-field tracking for low-altitude
  economy: An information-theoretic and data-driven framework,'' \emph{{IEEE}
  J. Sel. Topics Signal Process.}, pp. 1--17, 2026.

\bibitem{risleo}
R.~Deng \emph{et~al.}, ``Holographic {MIMO} for {LEO} satellite communications
  aided by reconfigurable holographic surfaces,'' \emph{{IEEE} J. Sel. Areas
  Commun.}, vol.~40, no.~10, pp. 3071--3085, Oct. 2022.

\bibitem{losris}
H.~Do \emph{et~al.}, ``Line-of-sight {MIMO} via intelligent reflecting
  surface,'' \emph{IEEE Trans. Wireless Commun.}, vol.~22, no.~6, pp.
  4215--4231, Jun. 2023.

\bibitem{risholonear}
Q.~Li \emph{et~al.}, ``Performance analysis of reconfigurable holographic
  surfaces in the near-field scenario of cell-free networks under hardware
  impairments,'' \emph{{IEEE} Trans. Wireless Commun.}, vol.~23, no.~9, pp.
  11\,972--11\,984, Sep. 2024.

\bibitem{risspa}
Z.~Li \emph{et~al.}, ``Performance analysis of {RIS}-assisted communications
  with element grouping and spatial correlation,'' vol.~12, no.~4, pp.
  630--634, Apr. 2023.

\bibitem{risexdof}
Y.~Cheng \emph{et~al.}, ``{RIS}-aided wireless communications: Extra degrees of
  freedom via rotation and location optimization,'' \emph{{IEEE} Trans.
  Wireless Commun.}, vol.~21, no.~8, pp. 6656--6671, Aug. 2022.

\bibitem{ristitdof}
H.~V. Cheng and W.~Yu, ``Degree-of-freedom of modulating information in the
  phases of reconfigurable intelligent surface,'' \emph{{IEEE} Trans. Inf.
  Theory}, vol.~70, no.~1, pp. 170--188, Jan. 2024.

\bibitem{CE4}
G.~Zhou \emph{et~al.}, ``Individual channel estimation for {RIS}-aided
  communication systems—a general framework,'' \emph{IEEE Trans. Wireless
  Commun.}, vol.~23, no.~9, pp. 12\,038--12\,053, Sep. 2024.

\bibitem{risphase1}
K.~Qi \emph{et~al.}, ``Reconfigurable-intelligent-surface-aided vehicular edge
  computing: Joint phase-shift optimization and multiuser power allocation,''
  \emph{IEEE Internet Things J.}, vol.~12, no.~1, pp. 764--777, Jan. 2025.

\bibitem{risphase2}
C.~B. Chaaya and M.~Bennis, ``{RIS} phase optimization via generative flow
  networks,'' \emph{IEEE Wireless Commun. Lett.}, vol.~13, no.~7, pp.
  1988--1992, Jul. 2024.

\bibitem{rismeasure}
S.~Kayraklik \emph{et~al.}, ``Indoor measurements for {RIS}-aided
  communication: Practical phase shift optimization, coverage enhancement, and
  physical layer security,'' \emph{IEEE Open J. Commun.}, vol.~5, pp.
  1243--1255, 2024.

\bibitem{rismeasure22}
J.~Wang \emph{et~al.}, ``Reconfigurable intelligent surface: Power consumption
  modeling and practical measurement validation,'' \emph{{IEEE} Trans.
  Commun.}, vol.~72, no.~9, pp. 5720--5734, Sep. 2024.

\bibitem{simpleris}
K.~Shafique and M.~Alhassoun, ``Going beyond a simple {RIS}: Trends and
  techniques paving the path of future {RIS},'' \emph{IEEE Open J. Antennas
  Propag.}, vol.~5, no.~2, pp. 256--276, Apr. 2024.

\bibitem{flexris}
M.~Yue \emph{et~al.}, ``{RIS}-assisted flexible receive spatial modulation for
  wireless communication,'' \emph{IEEE Wireless Commun. Lett.}, vol.~14, no.~7,
  pp. 1859--1863, Jul. 2025.

\bibitem{FAS}
K.-K. Wong \emph{et~al.}, ``Fluid antenna systems,'' \emph{{IEEE} Trans.
  Wireless Commun.}, vol.~20, no.~3, pp. 1950--1962, Mar. 2021.

\bibitem{fluidtut}
W.~K. New \emph{et~al.}, ``A tutorial on fluid antenna system for 6{G}
  networks: Encompassing communication theory, optimization methods and
  hardware designs,'' \emph{{IEEE} Commun. Surveys Tuts.}, vol.~27, no.~4, pp.
  2325--2377, Aug. 2025.

\bibitem{fasmag}
W.-J. Lu \emph{et~al.}, ``Fluid antennas: Reshaping intrinsic properties for
  flexible radiation characteristics in intelligent wireless networks,''
  \emph{IEEE Commun. Mag.}, vol.~63, no.~5, pp. 40--45, May 2025.

\bibitem{FASqout}
W.~K. New \emph{et~al.}, ``An information-theoretic characterization of
  {MIMO-FAS}: Optimization, diversity-multiplexing tradeoff and $q$-outage
  capacity,'' \emph{{IEEE} Trans. Wireless Commun.}, vol.~23, no.~6, pp.
  5541--5556, Jun. 2024.

\bibitem{perlimfas}
K.~K. Wong \emph{et~al.}, ``Performance limits of fluid antenna systems,''
  \emph{IEEE Commun. Lett.}, vol.~24, no.~11, pp. 2469--2472, Nov. 2020.

\bibitem{newcor}
P.~Ramírez-Espinosa \emph{et~al.}, ``A new spatial block-correlation model for
  fluid antenna systems,'' \emph{{IEEE} Trans. Wireless Commun.}, vol.~23,
  no.~11, pp. 15\,829--15\,843, Nov. 2024.

\bibitem{fasopdg}
W.~K. New \emph{et~al.}, ``Fluid antenna system: New insights on outage
  probability and diversity gain,'' \emph{{IEEE} Trans. Wireless Commun.},
  vol.~23, no.~1, pp. 128--140, Jan. 2024.

\bibitem{FRISlook}
A.~Salem \emph{et~al.}, ``On the performance enhancement potential of fluid
  reconfigurable intelligent surfaces,'' \emph{{IEEE} Trans. Wireless Commun.},
  vol.~25, pp. 14\,426--14\,439, 2026.

\bibitem{FRISonoff}
H.~Xiao \emph{et~al.}, ``Fluid reconfigurable intelligent surfaces:joint on-off
  selection and beamforming with discrete phase shifts,'' \emph{IEEE Wireless
  Commun. Lett.}, vol.~14, no.~10, pp. 3124--3128, Oct. 2025.

\bibitem{FRISpa}
F.~Ghadi \emph{et~al.}, ``Performance analysis of wireless communication
  systems assisted by fluid reconfigurable intelligent surfaces,'' \emph{IEEE
  Wireless Commun. Lett.}, vol.~14, no.~12, pp. 3922--3926, Dec. 2025.

\bibitem{FRISISAC}
J.~Ye \emph{et~al.}, ``Joint beamforming and position optimization for fluid
  {RIS}-aided {ISAC} systems,'' \emph{{IEEE} Trans. Commun.}, vol.~74, pp.
  11\,422--11\,438, 2026.

\bibitem{VegaSanchez2026}
J.~David Vega~Sánchez \emph{et~al.}, ``Understanding the role of phase and
  position design in fluid reconfigurable intelligent surfaces,'' \emph{IEEE
  Open J. Commun. Soc.}, vol.~7, pp. 1414--1425, 2026.

\bibitem{bvcfas}
T.~Wu \emph{et~al.}, ``Variable block-correlation modeling and optimization for
  secrecy analysis in fluid antenna systems,'' \emph{{IEEE} Trans. Wireless
  Commun.}, vol.~25, pp. 15\,069--15\,085, 2026.

\bibitem{pixfas}
J.~Zhang \emph{et~al.}, ``A novel pixel-based reconfigurable antenna applied in
  fluid antenna systems with high switching speed,'' \emph{IEEE Open J.
  Antennas Propag.}, vol.~6, no.~1, pp. 212--228, Feb. 2025.

\bibitem{trimimo}
M.~R. Castellanos \emph{et~al.}, ``Embracing reconfigurable antennas in the
  tri-hybrid {MIMO} architecture for {6G} and beyond,'' \emph{{IEEE} Trans.
  Commun.}, vol.~74, pp. 381--401, Jan. 2026.

\bibitem{trimimo22}
R.~W. Heath \emph{et~al.}, ``The tri-hybrid {MIMO} architecture,'' \emph{IEEE
  Wireless Commun.}, vol.~33, no.~1, pp. 199--206, Feb. 2026.

\bibitem{nfris}
D.~Jun \emph{et~al.}, ``Resource-efficient near-field misfocus mitigation in
  {RIS}-assisted wideband multi-user systems,'' \emph{IEEE Trans. Cogn. Commun.
  Netw.}, vol.~12, pp. 3148--3163, Jan. 2026.

\bibitem{fastnse}
H.~Hong \emph{et~al.}, ``A contemporary survey on fluid antenna systems:
  Fundamentals and networking perspectives,'' \emph{IEEE Trans. Netw. Sci.
  Eng}, vol.~13, pp. 2305--2328, 2026.

\bibitem{frissec}
X.~Zhu \emph{et~al.}, ``Fluid reconfigurable intelligent surface {(FRIS)}
  enabling secure wireless communications,'' \emph{IEEE Wireless Commun.
  Lett.}, vol.~15, pp. 2408--2412, 2026.

\bibitem{frispat}
H.~Xiao \emph{et~al.}, ``Fluid reconfigurable intelligent surface with
  element-level pattern reconfigurability: Beamforming and pattern co-design,''
  \emph{{IEEE} Trans. Wireless Commun.}, vol.~25, pp. 10\,791--10\,806, 2026.

\bibitem{FASRIS}
J.~Yao \emph{et~al.}, ``{FAS-RIS} communication: Model, analysis, and
  optimization,'' \emph{{IEEE} Trans. Veh. Technol.}, vol.~74, no.~6, pp.
  9938--9943, Jun. 2025.

\bibitem{nearris}
J.~Lee and S.~Hong, ``Near-field {LoS/NLoS} channel estimation for {RIS}-aided
  {MU-MIMO} systems: Piece-wise low-rank approximation approach,'' \emph{{IEEE}
  Trans. Wireless Commun.}, vol.~24, no.~6, pp. 4781--4792, Jun. 2025.

\bibitem{maxminris}
Q.-U.-A. Nadeem \emph{et~al.}, ``Asymptotic {Max-Min} {SINR} analysis of
  reconfigurable intelligent surface assisted {MISO} systems,'' \emph{{IEEE}
  Trans. Wireless Commun.}, vol.~19, no.~12, pp. 7748--7764, Dec. 2020.

\bibitem{FASRIS22}
J.~Yao \emph{et~al.}, ``A framework of {FAS-RIS} systems: Performance analysis
  and throughput optimization,'' \emph{{IEEE} Trans. Wireless Commun.},
  vol.~25, pp. 1333--1348, 2026.

\bibitem{neardai}
Y.~Lu and L.~Dai, ``Near-field channel estimation in mixed {LoS/NLoS}
  environments for extremely large-scale {MIMO} systems,'' \emph{{IEEE} Trans.
  Commun.}, vol.~71, no.~6, pp. 3694--3707, Jun. 2023.

\bibitem{marQ}
N.~Temme, ``Asymptotic and numerical aspects of the noncentral chi-square
  distribution,'' \emph{Comput. Math. Appl.}, vol.~25, no.~5, pp. 55--63, Mar.
  1993.

\bibitem{multio}
K.~Miettinen, \emph{Nonlinear Multiobjective Optimization}.\hskip 1em plus
  0.5em minus 0.4em\relax SpringerMcGraw-Hill, 1999.

\bibitem{fdris}
H.~Xiao \emph{et~al.}, ``Frequency diverse {RIS} {(FD-RIS)} enhanced wireless
  communications via joint distance-angle beamforming,'' \emph{IEEE Wireless
  Commun. Lett.}, vol.~13, no.~12, pp. 3459--3463, Dec. 2024.

\bibitem{CEM}
R.~Y. Rubinstein and D.~P. Kroese, \emph{The Cross-Entropy Method: A Unified
  Approach to Combinatorial Optimization, Monte-Carlo Simulation, and Machine
  Learning}.\hskip 1em plus 0.5em minus 0.4em\relax Springer, 2004.

\bibitem{sdr}
Z.-Q. Luo \emph{et~al.}, ``Semidefinite relaxation of quadratic optimization
  problems,'' \emph{{IEEE} Signal Process. Mag.}, vol.~27, no.~3, pp. 20--34,
  May 2010.

\bibitem{charnes}
P.~Zhu \emph{et~al.}, ``Integrated sensing and communication with
  reconfigurable holographic surface,'' \emph{{IEEE} Trans. Commun.}, vol.~73,
  no.~11, pp. 10\,377--10\,390, Nov. 2025.

\bibitem{boyd}
S.~Boyd and L.~Vandenberghe, \emph{Convex Optimization}.\hskip 1em plus 0.5em
  minus 0.4em\relax Cambridge, UK: Cambridge Univ. Press, 2004.

\bibitem{chol}
A.~Osinsky \emph{et~al.}, ``Regularization for cholesky decomposition in
  massive {MIMO} detection,'' \emph{{IEEE} Wireless Commun. Lett.}, vol.~12,
  no.~9, pp. 1603--1607, Sep. 2023.

\bibitem{Din}
A.~Zappone and E.~Jorswieck, ``Energy efficiency in wireless networks via
  fractional programming theory,'' \emph{Found. Trends Commun. Inf. Theory},
  vol.~11, no. 3-4, pp. 185--396, 2015.

\bibitem{Rieman1}
M.~AlaaEldin \emph{et~al.}, ``Optimization of energy-constrained {IRS-NOMA}
  using a complex circle manifold approach,'' \emph{IEEE Internet Things J.},
  vol.~11, no.~20, pp. 33\,133--33\,150, Oct. 2024.

\bibitem{Rieman2}
P.-A. Absil \emph{et~al.}, \emph{Optimization Algorithms on Matrix
  Manifolds}.\hskip 1em plus 0.5em minus 0.4em\relax Princeton University
  Press, 2008.

\bibitem{curse}
W.-Y. Fu, ``Accelerated high-dimensional global optimization: A particle swarm
  optimizer incorporating homogeneous learning and autophagy mechanisms,''
  \emph{Inf. Sci.}, vol. 648, p. 119573, Nov. 2023.

\bibitem{corrhalf}
D.-S. Shiu \emph{et~al.}, ``Fading correlation and its effect on the capacity
  of multielement antenna systems,'' \emph{{IEEE} Trans. Commun.}, vol.~48,
  no.~3, pp. 502--513, Mar. 2000.

\bibitem{bessbook}
M.~Abramowitz and I.~A. Stegun, \emph{Handbook of Mathematical Functions with
  Formulas, Graphs, and Mathematical Tables}.\hskip 1em plus 0.5em minus
  0.4em\relax National Bureau of Standards, 1964.

\end{thebibliography}
\end{document}